\documentclass[%
 reprint,
 superscriptaddress,
 amsmath,amssymb,
 aps,
prb,
]{revtex4-2}

\usepackage{graphicx}
\usepackage{subfigure}
\usepackage{float}
\usepackage{dcolumn}
\usepackage{bm}
\usepackage{rotating}
\usepackage{color}
\usepackage{ulem}
\usepackage{hyperref}

\begin{document}

\preprint{APS/123-QED}

\title{Extended imaginary gauge transformation in a general nonreciprocal lattice}

\author{Yunyao Qi}
    \affiliation{State Key Laboratory of Low-Dimensional Quantum Physics and Department of Physics, Tsinghua University, Beijing 100084, China}
\author{Jinghui Pi}
    \email{pijh14@gmail.com}
    \affiliation{State Key Laboratory of Low-Dimensional Quantum Physics and Department of Physics, Tsinghua University, Beijing 100084, China}
\author{Yuquan Wu}
    \affiliation{State Key Laboratory of Low-Dimensional Quantum Physics and Department of Physics, Tsinghua University, Beijing 100084, China}
\author{Heng Lin}
    \affiliation{State Key Laboratory of Low-Dimensional Quantum Physics and Department of Physics, Tsinghua University, Beijing 100084, China}
\author{Chao Zheng}
    \email{czheng@ncut.edu.cn}
    \affiliation{Department of Physics, College of Science, North China University of Technology, Beijing 100144, China}
\author{Gui-Lu Long}%
    \email{gllong@tsinghua.edu.cn}
    \affiliation{State Key Laboratory of Low-Dimensional Quantum Physics and Department of Physics, Tsinghua University, Beijing 100084, China}
    \affiliation{Beijing Academy of Quantum Information Sciences, Beijing 100193, China}
    \affiliation{Frontier Science Center for Quantum Information, Beijing 100084, China}
    \affiliation{Beijing National Research Center for Information Science and Technology, Beijing 100084, China}





\begin{abstract}
Imaginary gauge transformation (IGT) provides a clear understanding of the non-Hermitian skin effect by transforming the non-Hermitian Hamiltonians with real spectra into Hermitian ones. In this paper, we extend this approach to the complex spectrum regime in a general nonreciprocal lattice model. We unveil the validity of IGT hinges on a class of pseudo-Hermitian symmetry. The generalized Brillouin zone of Hamiltonians respect such pseudo-Hermiticity is demonstrated to be a circle, which enables easy access to the continuum bands, localization length of skin modes, and relevant topological numbers. Furthermore, we investigate the applicability of IGT and the underlying pseudo-Hermiticity beyond nearest-neighbor hopping, offering a graphical interpretation. Our theoretical framework is applied to establish bulk-boundary correspondence in the nonreciprocal trimer Su-Schrieffer-Heeger model and to analyze the localization behaviors of skin modes in the two-dimensional Hatano-Nelson model.
\end{abstract}

\maketitle
\section{\label{sec:Intro} Introduction}
Non-Hermitian physics has emerged as a rapidly growing field of study over the past few years \cite{Bender_2007, Peng_2014parity, Konotop_2016, Ashida_2018, Ashida_2020, Gopalakrishnan_2021}. The non-Hermiticity of the Hamiltonian arises when a system couples with its surroundings. Such systems encompass optical systems with gain and loss \cite{Feng_2017, Longhi_2017, El_2018, Ozawa_2019}, open systems with dissipation \cite{Rotter_2009}, and electron systems with finite-lifetime quasi-particles \cite{Yoshida_2018, Shen_2018, Yamamoto_2019}. A unique feature of the non-Hermitian system is the non-Hermitian skin effect (NHSE) \cite{Yao_2018, NHSE_2018}, namely the boundary localization of the majority of eigenstates. The existence of NHSE can lead to novel physical phenomena which have no Hermitian counterparts, including unidirectional physical effects \cite{Song_2019, Wanjura_2020, Xue_2022}, critical phenomena \cite{Li_2020, Liu_2020Helical, Yokomizo_2021, Guo_2021}, geometrical related effects in higher dimensions \cite{Sun_2021, Zhang_2022universal, Li_2022gain, Zhu_2022, Wu_2022complex} and so on. Experimental efforts to simulate non-Hermitian Hamiltonian and examine the corresponding physical effects have also made great progress \cite{Brandenbourger_2019non, Gao_2020anomalous, Ghatak_2020observation, Zhang_2021acoustic, Gao_2022non, Wang_2023extended, Helbig_2020, Hofmann_2020, Zou_2021, Weidemann_2020light, Song_2020twodimensional, Wang_2021braid, Kai_2021winding, Weidemann_2022topological, Li_2020topological, Liang_2022dynamic}. An important consequence of NHSE is the sensitivity of the spectra to the boundary conditions; for example, the open boundary spectra differ dramatically from the periodic boundary spectra \cite{Lee_2016}. In this case, the traditional bulk-boundary correspondence (BBC) no longer holds \cite{Ye_2018}. Alternative solutions to recover BBC with the existence of NHSE have become a main focus, and different approaches have been proposed \cite{Shen_2018topological, Yin_2018, Song_2019real, Kunst_2018, Yao_2018chern, GBZ_2019, Herviou_2019, Imura_2019, Imura_2020, Zirnstein_2021}. Among which the non-Bloch band theory \cite{Yao_2018, Yao_2018chern, GBZ_2019} provides a standard approach to deal with the non-negligible difference between periodic boundary conditions (PBCs) and open boundary conditions (OBCs) by introducing the concept of generalized Brillouin zone (GBZ). Systematic research on the topological modes and other novel effects in non-Hermitian systems has been conducted with the concept of GBZ \cite{Longhi_2019, Lee_2019anatomy, Kawabata_2020, Xiao_2020non, Lee_2020unraveling, Longhi_2020, Xue_2021simple, Wang_2021detecting, Fu_2022degeneracy, Wu_2022connections}. Moreover, NHSE itself has its topological origin \cite{kawabata_2019symmetry, Okuma_2020, Zhang_2020correspondence, Borgnia_2020}, which gives a different meaning of BBC and enriches the topological phases\cite{Gong_2018, Lee_2019correspondence, AGBZ_2020, Okuma_2023, Hu_2021knots, Lin_2023}.
 
On the other hand, the energy spectrum may be complex for a general non-Hermitian Hamiltonian. However, assuming the system exhibits $\eta$-pseudo-Hermitian symmetry, the eigenvalues are either real numbers or complex conjugate pairs \cite{Mostafazadeh_2002pseudo}. An example of such a system is a nonreciprocal lattice where all hopping matrix elements are real. An elegant method named imaginary gauge transformation (IGT) \cite{Hatano_1996, Hatano_1997} has been employed for specific nonreciprocal lattice models to connect non-Hermitian Hamiltonians under OBCs to their Hermitian counterparts when the energy spectra are purely real. This technique provides an intuitive framework for understanding the significant difference between the spectrum under OBC and PBC, as well as the existence of NHSE. In the simplest Hatano-Nelson (HN) model \cite{Hatano_1996}, IGT is employed to obtain the OBC spectrum and localization length of the skin modes. In the Su-Schrieffer-Heeger (SSH) model \cite{SSH_1979}, IGT helps to understand the breaking of conventional BBC with the existence of nonreciprocal hopping \cite{Yao_2018}. Other research utilizes the technique for certain models to shed light on the transition between real and complex spectrum \cite{Zeng_2022real}, and further extends it to the momentum space to address non-Hermiticity arising from complex potential \cite{img_2022}.

So far, most investigations involving IGT have been confined to the 1D nearest-neighbor (NN) hopping models within the real spectra regime [the shadowed area in Fig.~\ref{fig:1}(a)], where the non-Hermitian Hamiltonians can be transformed into their Hermitian counterparts. A comprehensive exploration of the relationship between the $\eta$-pseudo-Hermiticity and IGT in both real and complex spectra regimes remains to be undertaken. Additionally, while a generic nonreciprocal Hamiltonian with long-range hoppings cannot be transformed into a Hermitian counterpart even if its spectrum is entirely real \cite{Rafi_2024}, it is worth investigating the explicit condition that the IGT and relevant results are applicable with the existence of long-range hoppings. Considering the GBZ formalism as the standard approach in analyzing non-Hermitian Hamiltonians under OBCs, and the connection between IGT in real space and GBZ rescaling, the shape of GBZs in systems amenable to IGT is also of interest.

In this paper, we address the aforementioned questions and extend the results of IGT to more general nonreciprocal lattice Hamiltonians with complex spectra and long-range hoppings [the large circle in Fig.~\ref{fig:1}(a)]. We first elucidate the precise relationship between the IGT and pseudo-Hermiticity. While such nonreciprocal Hamiltonians are inherently $\eta-$pseudo-Hermitian, we demonstrate that the underlying reason for the applicability of IGT lies in the pseudo-Hermiticity characterized by a specific metric, namely $\eta_{\mathtt{I}}$ in this paper. Subsequently, we prove that the GBZ of such an $\eta_{\mathtt{I}}$-pseudo-Hermitian system is always a perfect circle in both real and complex spectra regimes, which cannot be directly obtained with IGT. Furthermore, we establish the sufficient and necessary condition of $\eta_{\mathtt{I}}$-pseudo-Hermiticity in the presence of long-range hopping terms. This condition can be interpreted as a simple picture that the product of asymmetric ratios between any two sites should be path-independent. Leveraging this condition, we can effortlessly extend the IGT technique to certain two-dimensional (2D) cases.

The key insight of this paper is that $\eta_{\mathtt{I}}$ is valid in both the symmetry exact phase and symmetry broken phase, suggesting the presence of shared characteristics across these phases. From a detailed analysis of the characteristic equation, we summarize these characteristics as the circular GBZ. The characteristic of circular GBZ can even be generalized to systems with complex hoppings, where $\eta$-pseudo-Hermiticity no longer holds. This allows for the parametrizing of GBZ with radius $r$ as $\beta = re^{ik}$, enabling the calculation of the continuum band spectrum, wave function, and relevant topological numbers using the same approach employed for the Hermitian case, where similar calculations are performed in the Brillouin Zone (BZ) $\beta=e^{ik}$. The important aspect of our paper lies in the applicability of this result in both phases and Hamiltonians with complex hoppings, ensuring the effectiveness of the procedure even when the spectrum is complex and the system cannot be transformed into a Hermitian counterpart via IGT.

The rest of the paper is organized as follows: In Sec.~\ref{sec:Eta}, we first give an overview of the NHSE and how it can be understood from IGT. Then we introduce the theory of $\eta$-pseudo-Hermiticity \cite{Mostafazadeh_2002pseudo} and derive the relation between IGT and the $\eta_{\mathtt{I}}$ metric. We give a detailed discussion of the behavior of $\eta_{\mathtt{I}}$ in both symmetry exact and broken phase. In Sec.~\ref{sec:GBZ}, we prove the GBZ of $\eta_{\mathtt{I}}$-pseudo-Hermitian Hamiltonian is a perfect circle, with the radius only relevant to the modulus of hopping strength. In Sec.~\ref{sec:NNN}, we derive the general condition for $\eta_{\mathtt{I}}$-pseudo-Hermiticity, which extends the IGT to nonreciprocal lattices with long-range hopping terms. In Sec.~\ref{sec:app}, we apply our theoretical results to establish the BBC for non-Hermitian trimer SSH model and obtain the NHSE of 2D HN model.

\section{\label{sec:Eta} Imaginary gauge transformation and $\eta$-pseudo-Hermiticity}
\subsection{\label{sec:NHSE}Imaginary gauge transformation and NHSE}

\begin{figure}[t]
    \includegraphics[width=\columnwidth]{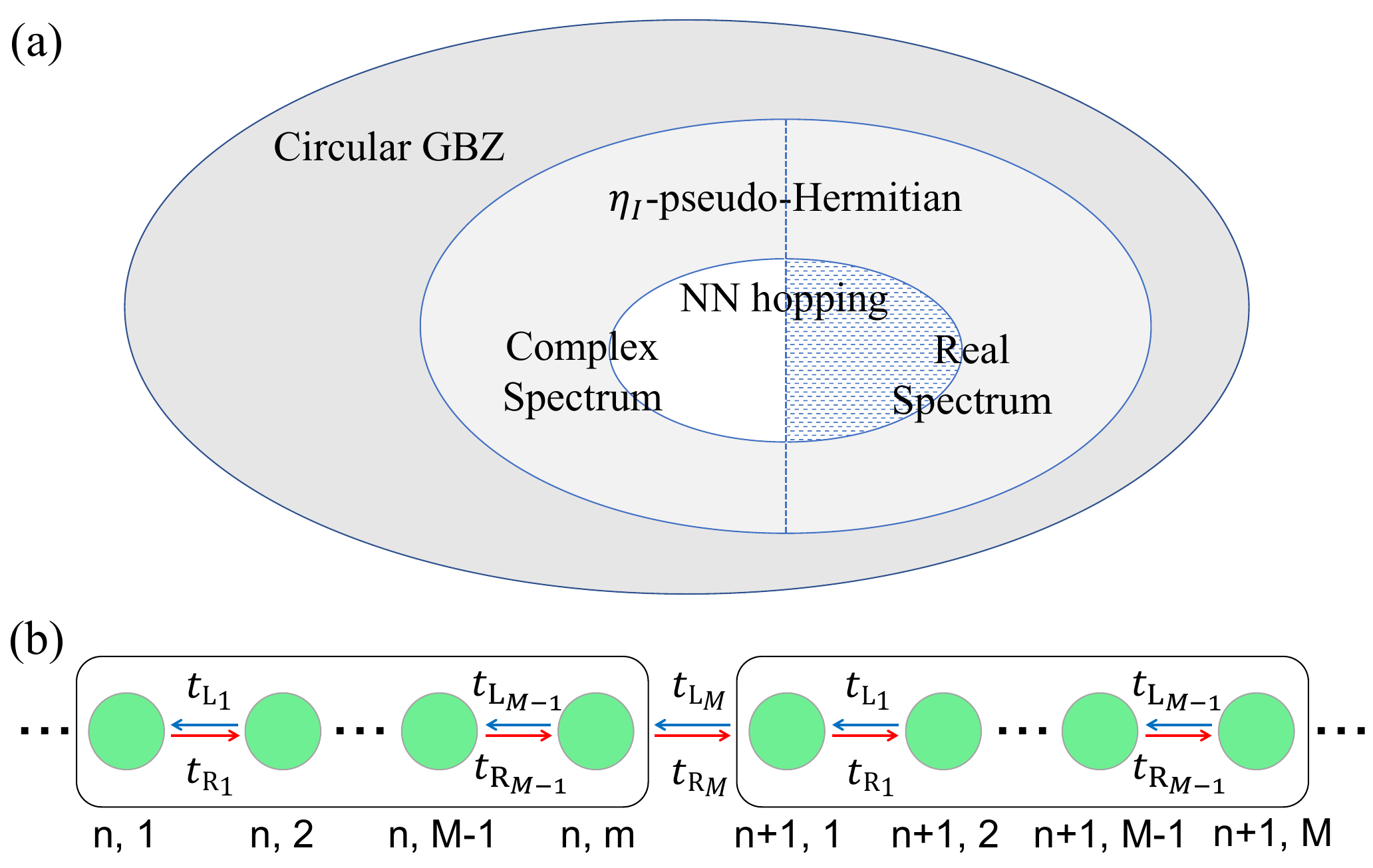}
  \caption{(a) The logical relationship between the main concepts in this paper. The application of IGT in previous paper is limited to the shadowed area, which stands for the NN hopping models with entirely real spectra. With the help of $\eta_{\mathtt{I}}$-pseudo-Hermiticity, we extend it to the complex spectra regime and beyond NN hopping. We further demonstrate that it can also be extended to complex hopping models since the GBZ of the system is circular. (b) The generic nonreciprocal model with NN hoppings.} 
  \label{fig:1}
\end{figure}

The IGT provides an intuitive way to understand the NHSE. A simple example of applying the IGT to HN models with real spectra to obtain the NHSE can be found in Appendix~\ref{app:hn}. In this section, we briefly review the IGT generalized to nonreciprocal lattices with more than one sublattice in a unit cell as shown in Fig.\ref{fig:1}(b) \cite{Yao_2018, Zeng_2022real}. Consider a general one-dimensional (1D) OBC nonreciprocal lattice with $N$ unit cells and $M$ sublattices in each unit cell. The Hamiltonian with only NN hoppings is given by
\begin{equation}
\label{eq:NN_def}
\begin{aligned}
H_\mathtt{NN}  & =\sum_{n=1}^{N}  \sum_{i=1}^{M-1}(t_{\mathtt{R}_{i}} a_{n,i+1}^{\dagger
} a_{n,i}+t_{\mathtt{L}_{i}} a_{n,i}^{\dagger} a_{n,i+1})\\
& +\sum_{n=1}^{N-1}(t_{\mathtt{R}_{M}}a_{n+1,1}^{\dagger}a_{n,M}+t_{\mathtt{L}_{M}}a_{n,M}^{\dagger}a_{n+1,1}),
\end{aligned}
\end{equation}
where $a_{n,i}^{\dagger}$ ($a_{n,i}$) are the creation (annihilation) operators for the $i$-th sublattice in the $n$-th unit cell. ${t_{\mathtt{R}_{i}/\mathtt{L}_{i}}\in\mathbb{R}, i=1,\cdots,M}$ are the hopping amplitudes and $i=M$ ($i\neq M$) stand for the intercell (intracell) hopping. If  $t_{\mathtt{L}_{i}} t_{\mathtt{R}_{i}}> 0$ for all hopping amplitudes, $H_\mathtt{NN}$ can be related to a Hermitian Hamiltonian $H^{\prime}_\mathtt{NN}$ via an IGT, which is given by the following diagonal matrix
\begin{equation}
  S_{\mathtt{NN}} = \mathtt{diag}\{r_M, r_M^2,\cdots,r_M^N\} \otimes \mathtt{diag}\{r_0, r_1, \cdots, r_{M-1}\}
    \label{eq:transform_def}
\end{equation}
with $r_{i} = \sqrt{\frac{t_{\mathtt{R}_{1}}\cdots
t_{\mathtt{R}_{i}}}{t_{\mathtt{L}_{1}}\cdots t_{\mathtt{L}_{i}} }}$ for $i=1,\cdots,M$ and $r_0=1$, or symbolically, 
\begin{equation}
H^{\prime}_\mathtt{NN}=S^{-1}_\mathtt{NN} H_\mathtt{NN} S_\mathtt{NN}.
\end{equation}

Although $H^{\prime}_\mathtt{NN}$ and $H_\mathtt{NN}$ share the same spectrum, their eigenstates exhibit distinct localization behavior. The bulk eigenstates of $H^{\prime}_\mathtt{NN}$ are extended because of Bloch's theorem, while the majority eigenstates of $H_\mathtt{NN}$ are localized at the boundary when $r_{M} \neq 1$, namely they feature the NHSE. Besides, the NHSE of $H_\mathtt{NN}$ is determined by $r_{M}$. All skin modes are localized at the left (right) edge when $r_M <1$  $(r_M >1)$ with the same localization length $\left\vert \ln r_{M}\right\vert ^{-1}$. While the localization length for different skin modes is generally different, the system that IGT is applicable has a unified localization length for all skin modes because every bulk state in the Hermitian counterpart is modulated by the same exponential envelope.

\subsection{\label{sec:pseudo} $\eta$-pseudo-Hermiticity and imaginary gauge transformation }

The aforementioned IGT is closely connected to the $\eta$-pseudo-Hermiticity of Hamiltonian. The definition of the $\eta$-pseudo-Hermitian \cite{Mostafazadeh_2002pseudo} $H$ is that there exists an invertible Hermitian operator $\eta$ satisfies 
\begin{equation}
    \eta H = H^{\dagger} \eta.
    \label{eq:eta_def}
\end{equation}
This definition can be regarded as a generalization of Hermiticity since the $\eta$-pseudo-Hermitian Hamiltonian is self-adjoint under the inner product defined by $\eta$. We provide a brief explanation of the main conclusions in $\eta$-pseudo-Hermitian Hamiltonians proposed by \cite{Mostafazadeh_2002pseudo} in Appendix~\ref{app:pseudo}. In this section, we utilize these conclusions to draw an exact map between $\eta$-pseudo-Hermiticity and IGT. While $\eta$-pseudo-Hermitian Hamiltonians have similar properties as Hermitian ones, they possess some unique properties \cite{Mostafazadeh_2002pseudo}. The most important one is the eigenvalues come in either real values or complex conjugate pairs (this is a necessary and sufficient condition for $\eta$-pseudo-Hermiticity), which can be understood from the fact that the inner product defined by $\eta$ can be either positive definite or indefinite. There may also exist more than one distinct $\eta$ operator for a pseudo-Hermitian Hamiltonian with certain symmetry (see Appendix~\ref{app:reflection-symmetry} for an example). 

Since the orthogonality of eigenstates is no longer guaranteed because of the non-Hermiticity, the bi-orthonormal basis is typically employed. This basis comprises the right eigenstates $ | \psi_i \rangle$ and left eigenstates $|\phi_i \rangle$, satisfying the eigen-equations
\begin{equation}
    H | \psi_i \rangle = E_i | \psi_i \rangle, \quad H^{\dagger} | \phi_i \rangle =E_i^* | \phi_i \rangle,
\end{equation}
with the completeness and bi-orthogonal relations
\begin{equation}
     \quad \sum_i | \psi_i \rangle \langle \phi_i | = 1, \quad \langle \psi_i | \phi_j \rangle = \delta_{ij}.
      \label{eq:bi-orthonormal}
\end{equation}
These conditions hold when the system is not at the exceptional points (EPs), where eigenstates coalesce \cite{Heiss_2012}. In the following discussion, we focus on the cases where 
Eq.~(\ref{eq:bi-orthonormal}) holds.  

We shall focus on the positive definite $\eta$ first. Any positive definite $\eta_{\mathtt{P}}$ operator can be decomposed as 
\begin{equation} \label{eq:decompose}
    \eta_{\mathtt{P}} = \Sigma^{\dagger}\Sigma,
\end{equation}
where $\Sigma$ can be regarded as the "square root" of $\eta_{\mathtt{P}}$. By introducing the metric operator $\Sigma$, the $\eta$-pseudo-Hermitian Hamiltonian $H$ can be transformed into a Hermitian Hamiltonian $H^{\prime}$  through the relation $H^{\prime} = \Sigma H \Sigma^{-1}$. Let the orthonormal basis of $H^{\prime}$ be $\{|\psi^{\prime}_i\rangle\}_{i=1,\cdots, N}$. Then the bi-orthonormal basis of $H$ and $H^{\dagger}$ can be constructed as $|\psi_i\rangle=\Sigma^{-1}|\psi^{\prime}_i\rangle, |\phi_i\rangle=\Sigma^{\dagger}|\psi^{\prime}_i\rangle$. The completeness of $\{|\psi^{\prime}_i\rangle\}_{i=1,\cdots, N}$ leads to 
\begin{equation}\label{eq:positive}
    \eta_{\mathtt{P}}=\sum_i |\phi_i\rangle\langle\phi_i|.
\end{equation}
 Reference \cite{Zeng_2022} numerically verified this formula in a NN hopping lattice in the real spectrum regime to show that it is $\eta$-pseudo-Hermitian. Here we can give a clear theoretical explanation that the aforementioned IGT is just the inverse matrix of $\Sigma$ in Eq.~(\ref{eq:decompose}). More specifically, the positive definite metric generated by the IGT is given by
\begin{equation}
\label{eta_I}
\begin{aligned}
    \eta_{\mathtt{I}} = & S_\mathtt{NN}^{-2}\\
    = & \mathtt{diag}\{R_m, R_m^2,\cdots,R_M^N\}\otimes \\
    &\mathtt{diag}\{R_0, R_1,\cdots,R_{M-1}\},
\end{aligned}
\end{equation}
with $R_i = r_i^{-2}=\frac{t_{\mathtt{L}_{1}}\cdots
t_{\mathtt{L}_{i}}}{t_{\mathtt{R}_{1}}\cdots t_{\mathtt{R}_{i}}}$ for $i=1,\cdots,M$ and $R_0=1$. Such $\eta_{\mathtt{I}}$ has a diagonal form, providing an intuitive understanding of its effect. The Hamiltonian becomes "Hermitian" after a simple rescaling of the inner-product space, and the exponentially growing weight repels the distribution of eigenstates to the other edge. Figures ~\ref{fig:NHSE}(a)--(c) show an example of the NHSE of NN hopping nonreciprocal lattice with three sublattices in one unit cell in the PT-exact phase. The energy spectrum is purely real, as the IGT can map it to its Hermitian counterpart. The majority of the eigenstates exhibit exponential decay from the edge with the same decay rate, which is well predicted by the theoretical envelope. Note that four isolated energy levels are away from the continuum bands, which correspond to the conventional topological boundary states \cite{AGBZ_2020}. We only focus on the non-Hermitian skin modes in this section, and the conventional 
topological boundary states are not shown in the plot of eigenstates. In Sec.~\ref{sec:SSH3}, we will discuss those conventional topological boundary states in detail.

Before going to the indefinite $\eta$ case, we remember that PT-symmetric Hamiltonians \cite{Bender_98PT} are all $\eta$-pseudo-Hermitian \cite{Mostafazadeh_2002pseudo}. Since the PT operator can be generalized to any anti-unitary operator \cite{PT_2002}, the general nonreciprocal Hamiltonians with real hopping terms exhibit the generalized PT symmetry because they are invariant under the complex conjugate operation, or symbolically $H=KHK$. We will use the PT-exact phase and PT-broken phase to distinguish the real or complex spectrum in the following discussion.

\subsection{\label{sec:PT-broken} $\eta_{\mathtt{I}}$-pseudo-Hermiticity in the PT-broken phase}

 The procedure in Sec.~\ref{sec:NHSE} is only valid in the PT-exact phase, where $\eta_{\mathtt{I}}$ is positive definite. More precisely, for a general Hamiltonian with NN hopping defined in Eq.~(\ref{eq:NN_def}), the new Hamiltonian $H^{\prime}_\mathtt{NN}$ generated by the IGT is Hermitian, and $\eta_{\mathtt{I}}$ is positive definite provided that $t_{\mathtt{R}_i}t_{\mathtt{L}_i} > 0$ holds for all $i=1,2,\cdots,M$. However, the fact that $\eta_{\mathtt{I}}$ is not necessarily positive definite indicates that $\eta_{\mathtt{I}}$-pseudo-Hermiticity can be used to explore the NHSE in the PT-broken phase. Although no Hermitian counterpart exists in the PT-broken phase, the $\eta_{\mathtt{I}}$-pseudo-Hermitian symmetry is still respected.  The only difference is that $\eta_{\mathtt{I}}$ has negative eigenvalues, i.e., there exists some $i$ such that $R_i < 0$. In this case, the $\eta_{\mathtt{I}}$ operator is indefinite and Eq.~(\ref{eq:positive}) is not applicable.
 
 However, we can find a general expression of the $\eta$ operator no matter if it is positive definite or not. We use $i_+$ and $i_-$ to denote the eigenstates with complex conjugate eigenvalues, while $i_0$ represents eigenstates with real eigenvalues. If the energy spectrum is non-degenerate, we have $\eta|\psi_{i_+}\rangle$ is proportional to $|\phi_{i_-} \rangle$ and vice versa. This can be expressed as \cite{Mostafazadeh_2002pseudo}
\begin{equation}\label{eq:coeff}
    \eta|\psi_{i_{\pm}}\rangle = c_{i_{\pm}}|\phi_{i_{\mp}}\rangle, \quad \eta|\psi_{i_{0}}\rangle = c_{i_{0}}|\phi_{i_{0}}\rangle,
\end{equation}
where $c_{i_{\pm}}$ and $c_{i_{0}}$ are the proportional coefficients. We can diagonalize the superposition coefficients in the characteristic subspace for the degenerate energy spectrum to get the above relation . If all $c_{i_0} \in \mathbb{R}^+$, we can simultaneously adjust the normalization coefficients of $|\psi\rangle$ and $|\phi\rangle$ to ensure that all coefficients $c_{i_{\pm}}$ and $c_{i_{0}}$ are equal to one while preserving the bi-orthonormal condition. Otherwise, some $c_{i_0}$ may take the value $-1$ (see Appendix~\ref{app:eta}). Thus, by substituting Eq.~(\ref{eq:coeff}) into the completeness of bi-orthonormal basis described by Eq.~(\ref{eq:bi-orthonormal}), we can express the $\eta$ operator formally as
\begin{equation} \label{eq:eta_F}
    \eta = \sum_{i_{\pm}} \left\vert \phi_{i_{\pm}} \right\rangle \left\langle \phi_{i_{\mp}}\right\vert + \sum_{i_0}c_{i_0}\left\vert \phi_{i_{0}} \right\rangle \left\langle \phi_{i_{0}}\right\vert,
\end{equation} 
where $c_{i_0}$ can take the values $\pm 1$. Note that Eq.~(\ref{eq:eta_F}) allows for the construction of different $\eta$ operators by varying the $i$-dependent normalization coefficients in the substitution $|\phi_i\rangle \rightarrow a_i|\phi_i\rangle$. Thus, the $\eta$ operator is not unique for a given Hamiltonian. While a positive definite $\eta$ operator ensures an entirely real spectrum, as shown in Sec.~\ref{sec:pseudo}, an indefinite $\eta$ operator does not guarantee the presence of complex eigenvalues. An example is shown in Appendix~\ref{app:reflection-symmetry}. The reason is that even for an entirely real spectrum, it is possible to construct an indefinite $\eta$ operator using Eq.~(\ref{eq:eta_F}) by setting  $c_{i_0}=-1$ for some $i_0$. The $\eta_{\mathtt{I}}$ operator we discuss here is positive definite in the PT-exact phase and indefinite in the PT-broken phase. 
 
\begin{figure*}
	\centering
	 \includegraphics[width=0.95\textwidth]{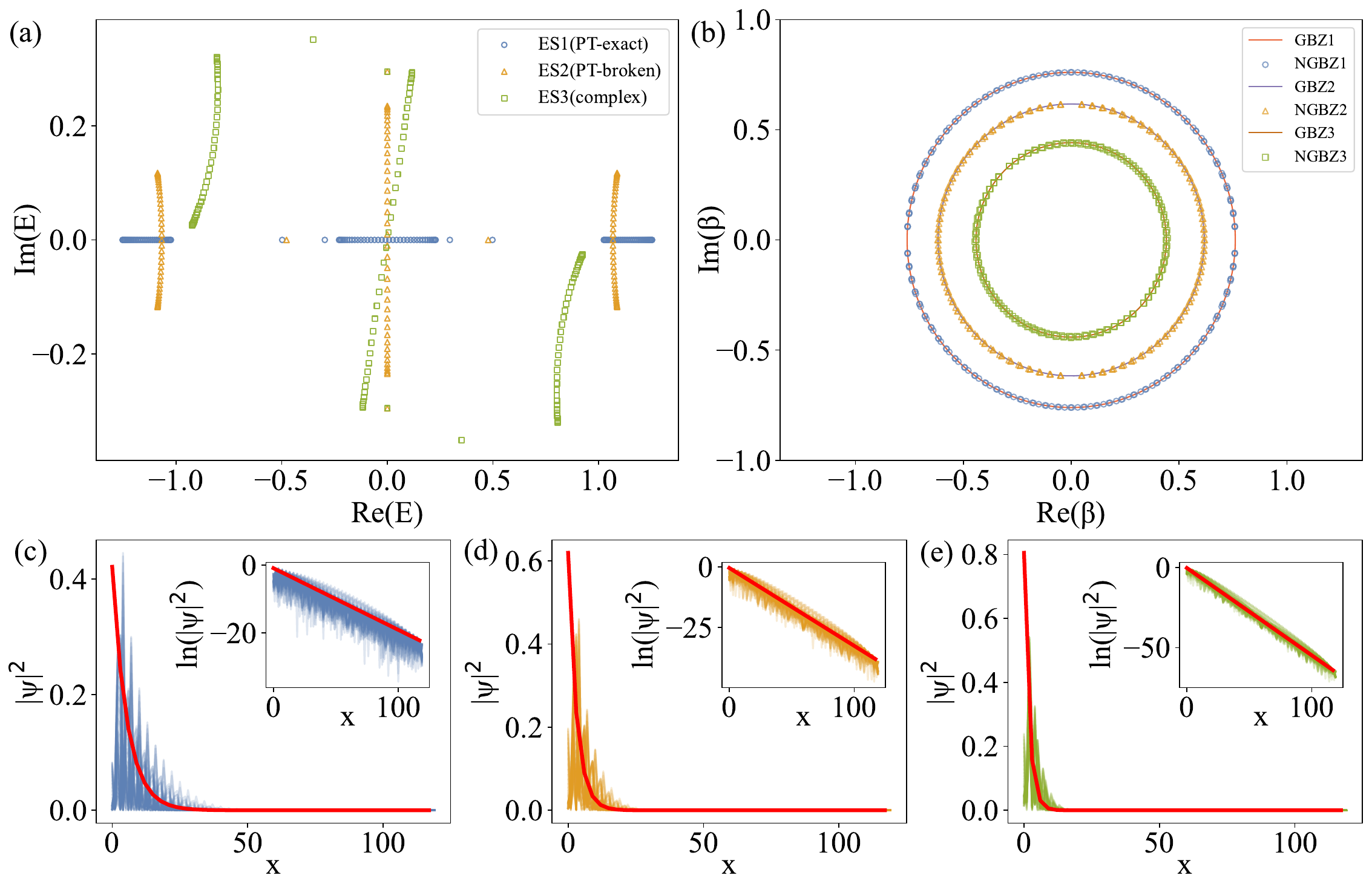}
    	\caption{ (a) The energy spectra of SSH3 model in the PT-exact phase, the PT-broken phase, and with complex hoppings. The system size is $N=40$. Apart from the continuum band spectrum composed of the majority of eigenvalues, some isolated discrete energy levels exist. (b) The GBZs correspond to the energy spectra in (a), which are obtained by numerically solving the characteristic equations (discrete points) and analytically calculating the assisted GBZs (solid lines) \cite{AGBZ_2020}. The shape of each GBZ is shown to be a circle and the radius is only relevant to the modulus of hopping strength. The corresponding squared modulus of continuum band eigenstates in the PT-exact phase (c), the PT-broken phase (d), and with complex hoppings (e); the red lines represent the theoretical exponential envelopes. Inset plots are the results on a logarithmic scale.  
}
	\label{fig:NHSE}
\end{figure*}

 The unified expression for $\eta_{\mathtt{I}}$ in the two phases also indicates that the NHSE in the PT-broken phase, which cannot be straightforwardly shown by the IGT, may have the same behavior as in the PT-exact phase. As shown in Fig.~\ref{fig:NHSE}(d), all skin modes share the same decay behavior, more precisely, the same localization length, in the PT-broken phases, just like in the PT-exact phase shown in Fig.~\ref{fig:NHSE}(c), although the energy spectrum has significant differences.

\section{\label{sec:GBZ} Analysis from the GBZ}
The non-Bloch band theory \cite{Imura_2019, GBZ_2019, Imura_2020} provides a systematic approach to analyzing the spatial periodic non-Hermitian system under OBC utilizing the GBZ (see Appendix~\ref{app:GBZ}). In this section, we demonstrate the unified property in both PT-exact and PT-broken phases using the concept of GBZ. The real space Hamiltonian of a tight-binding model can be expressed as the block form 
\begin{equation} \label{eq:tb}
    H_{\mathtt{TB}} = \sum_n \sum_{m=-p}^{p}\sum_{ij} T^{ij}_{m} a^{\dagger}_{n+m, i}a_{n, j},
\end{equation}
where $p$ is the longest hopping range related to the unit cell. The matrix $T_m^{ij}$ describes the hopping strength between the $i$th and $j$th sublattice of $n$th and $(n+m)$th unit cell. In the non-Bloch theory, the generalization of the wave vector is taken as $e^{ik}\rightarrow \beta$ \cite{GBZ_2019}. Then, analogous to the bulk Hamiltonians $H(k)$ in Hermitian cases, the generalized Bloch Hamiltonian $H(\beta)$ can be formally expressed as 
\begin{equation} \label{eq:beta}
    H_{\mathtt{TB}}(\beta) = \sum_{m=-p}^{p} T_m \beta^{-m}.
\end{equation}
The fundamental result of non-Bloch theory for 1D non-Hermitian systems without symmetry is that the GBZ is defined by the condition $|\beta_p|=|\beta_{p+1}|$, where $\beta_{p}$ and $\beta_{p+1}$ represent the $p$th and $(p+1)$th $\beta$ values respectively, when sorted in ascending order of magnitude by $|\beta_1| \leq |\beta_2| \leq \cdots \leq |\beta_{2p}|$ \cite{GBZ_2019}. The GBZ is formed by tracing the trajectory of these $\beta_{p}$ and $\beta_{p+1}$ values across different energy levels within the continuum bands.

Here we show that the GBZ of the aforementioned $\eta_{\mathtt{I}}$-pseudo-Hermitian Hamiltonian is indeed a circle. To establish this, we first show that the similarity transformation defined by the matrix 
\begin{equation}
    S(a, A) =\mathtt{diag}\{a,a^{2},\cdots,a^{N}\}\otimes A,
\end{equation}
where $A$ is arbitrary invertible matrix and $a\in\mathbb{C}$, can transform every $\beta$ to $\beta/a$ of a tight-binding Hamiltonian given by Eq.~(\ref{eq:tb}). This transformation can be interpreted as a rescaling of the GBZ. The proof is straightforward
\begin{equation}
\begin{aligned}
    H^{\prime}_{\mathtt{tb}} &= S(a,A)^{-1}H_{\mathtt{tb}}S(a,A) \\
    &=\sum_n \sum_{m=-P}^{P}\sum_{ij} a^{-m} (A^{-1}T_{m}A)^{ij} a^{\dagger}_{n+m, i}a_{n, j},
\end{aligned}
\end{equation}
which implies that the hopping matrix is transformed as $T_m \rightarrow a^{-m}A^{-1}T_mA$. Thus, the transformation for $H_{\mathtt{tb}}(\beta)$ is
\begin{equation}
    H_{\mathtt{tb}}^{\prime}(\beta)= A^{-1}(\sum_{m=-P}^{P}T_m(a\beta)^{-m})A.
\end{equation}
Since eigenvalues are invariant under similarity transformation, the $\beta$ in the GBZ of $H_{\mathtt{tb}}$ corresponds to $a\beta$ of $H_{\mathtt{tb}}^{\prime}$. Hence, the transformation for GBZ is $\beta \rightarrow \beta/a$. 

We note that the IGT in Eq.~(\ref{eq:transform_def}) can be represented as $S(r_{M}, \rm{diag}\{r_0, \cdots, r_{M-1}\})$. In the PT-exact phase, the IGT can transform the Hamiltonian into a Hermitian matrix. The GBZ after the transformation is the unit circle because of Hermiticity, so the GBZ of the original Hamiltonian is a circle with radius 
\begin{equation}
    |\beta_\mathtt{PT}| = r_M = \sqrt{\frac{t_{\mathtt{R}_1}\cdots t_{\mathtt{R}_M}}{t_{\mathtt{L}_1}\cdots t_{\mathtt{L}_M}}}.
\end{equation} 
The question arises as to whether this circular GBZ persists in the PT-broken phase. To address this, we first present an intuitive analysis from the perspective of real space, followed by a rigorous proof in the $\beta$ space. Notably, the $\eta_{\mathtt{I}}$ operator can be expressed as $S(r_M^{-2},\rm{diag}\{r_0^{-2},\cdots,r_{M-1}^{-2}\})$. This implies that the GBZ for $H_{\mathtt{tb}}$ and $H^{\dagger}_{\mathtt{tb}}$ are linked by the transformation $\beta \leftrightarrow \beta/r_M^2$. 
Furthermore, utilizing Eqs.~(\ref{eq:tb}) and (\ref{eq:beta}), we derive the generalized Bloch Hamiltonian for $H^{\dagger}_{\mathtt{tb}}$ as
\begin{equation}
    H^{\dagger}_{\mathtt{tb}}(\beta) = \sum_{m=-p}^{p} T_m^{\dagger}\beta^{m}.
\end{equation}
Since the eigenvalues of $H_{\mathtt{tb}}$ and $H_{\mathtt{tb}}^{\dagger}$ are complex conjugate pairs, the solutions for $\beta$ that satisfy the characteristic equation $|H_{\mathtt{tb}}(\beta)-E|=0$ and $|H_{\mathtt{tb}}^{\dagger}(\beta)-E^*|=0$ are related by the transformation $\beta \leftrightarrow 1/\beta^*$. This relationship, along with $\beta \leftrightarrow \beta/r_M^2$ resulting from $\eta_{\mathtt{I}}$, suggests that the GBZ for generic Hamiltonians which respect $\eta_{\mathtt{I}}$-pseudo-Hermitian symmetry should be a circle with radius 
\begin{equation} \label{eq:GBZ}
    |\beta| = |r_M| = \sqrt{|\frac{t_{\mathtt{R}_1}\cdots t_{\mathtt{R}_M}}{t_{\mathtt{L}_1}\cdots t_{\mathtt{L}_M}}|}.
\end{equation}
This holds regardless of whether the PT-symmetry is broken or not. To rigorously prove Eq.~(\ref{eq:GBZ}) in both the PT-exact and PT-broken phase, we can utilize the straightforward relation given by the similarity transformation of the generalized Bloch Hamiltonian
\begin{equation}\label{eq:H(beta)_transform}
   S_{\eta}^{-1} H(r_M^2 \beta^{-1}) S_{\eta} = H^{\mathtt{T}}(\beta),
\end{equation}
where $H^{\mathtt{T}}(\beta)$ denotes the transpose of $H(\beta)$, and 
\begin{equation}
    S_{\eta}=\mathtt{diag}\{1, r_1^2, \cdots, r_{M-1}^2\}
\end{equation}
is just the second part of Eq.~(\ref{eta_I}). Since $H^{\mathtt{T}}(\beta)$ shares the same spectrum with $H(\beta)$, and that the spectrum remains invariant under similarity transformation, we can conclude that for every solution for $\beta$ that satisfies $|H(\beta)-E|=0$, $r_M^2\beta^{-1}$, satisfies $|H(r_M^2\beta^{-1})-E|=0$. In other words, the solution for $\beta$ forms pairs $(\beta, r_M^2 \beta^{-1})$. Combining this with the condition for GBZ $|\beta_p|=|\beta_{p+1}|$, the GBZ is determined by $|\beta| = |r_M|$, leading to Eq.~(\ref{eq:GBZ}) in both phases. This result rigorously demonstrates the duality of NHSE in the PT-broken and the PT-exact phases. Since the GBZ is a circle with a radius solely dependent on the absolute value of hopping strengths, every Hamiltonian in the PT-broken phase has a counterpart in the PT-exact phase sharing the same GBZ. Notably, the localization length of the skin modes equals $|\ln|\beta||^{-1}$, implying that the localization behavior of the skin modes in the PT-broken phase mirrors their counterparts in the PT-exact phase. We emphasize that the invertibility is the only requirement for $S_{\eta}$ in the above derivation. Even when the complex hopping breaks the $\eta_{\mathtt{I}}$-pseudo-Hermiticity, the above derivation is still valid. Thus, the conclusion can be generalized to the case $t_{\mathtt{L}_i/\mathtt{R}_i} \in \mathbb{C}$.

The circular GBZ deforms to abnormal shapes with zero or infinite radius at EPs, which marks the transition between the PT-exact and PT-broken phases. This can be understood from the fact that the geometric origin of EPs is the existence of cusps in the GBZ \cite{Hu_2024}. Since there is no cusp in a circular GBZ, the EPs only exist in abnormal cases with zero or infinite radius. At EPs, at least one hopping term reaches zero. The IGT and corresponding $\eta_{\mathtt{I}}$ operator deforms to zero or infinite matrix. Therefore, the IGT cannot deal with systems with EPs.

In Fig.~\ref{fig:NHSE}, we verify our conclusion by numerically obtaining the spectra, GBZs, and the distribution of continuum band states in different parameter regimes. We can see that the GBZs are always circular in Fig.~\ref{fig:NHSE}(b), and all skin modes exhibit the same localization length in Figs.~\ref{fig:NHSE}(c)--\ref{fig:NHSE}(e). This result transcends the limitations of the real spectrum inherent in the conventional IGT approach. The simple shape of the GBZ facilitates the determination of the continuum bands in the thermodynamic limit and the relevant topological numbers defined in the GBZs.

\section{\label{sec:NNN} Imaginary gauge transformation beyond NN hopping}
Previous paper has only applied IGT to Hamiltonians with NN hoppings, as other hopping terms with longer ranges cannot be effectively balanced together with the NN hoppings in general. In this part, we explore the condition under which IGT remains valid in the presence of long-range hoppings, enabling its application to more complex scenarios.

Without loss of generality, we introduce long-range nonreciprocal hoppings between the $i$th sublattice of the $n$th unit cell and the $j$th sublattice of ($n+m$)th unit cell with hopping strengths $t_\mathtt{R}^{\prime}$ and $t_\mathtt{L}^{\prime}$, respectively (the concept NN in this paper refers to the nearest sublattice, not the nearest unit cell). Then, the Hamiltonian can be expressed as
\begin{equation}
    H_{\mathtt{Long}} = H_{\mathtt{NN}} + \sum_{n}(t_\mathtt{R}^{\prime}a^{\dagger}_{n+m,j}a_{n,i}     +t_\mathtt{L}^{\prime}a^{\dagger}_{n,i}a_{n+m,j}).
\end{equation}
 The effect of IGT on the creation and annihilation operators is given by
\begin{equation}
    c_{n,i}^{\dagger} = r_{i-1}r_M^n a_{n,i}^{\dagger}, \quad c_{n,i} = r_{i-1}^{-1}r_M^{-n} a_{n,i}.
\end{equation}
For a Hermitian counterpart of $H_{\mathtt{Long}}$ (in the PT-exact phase) to exist, the IGT must satisfy the necessary and sufficient condition:
\begin{equation} \label{eq:NNN}
    \frac{t_\mathtt{R}^{\prime}}{t_\mathtt{L}^{\prime}} = \left(\frac{r_{j-1}r_M^m}{r_{i-1}}\right)^2.
\end{equation}
Furthermore, it serves as a necessary and sufficient condition for the underlying $\eta_{\mathtt{I}}$-pseudo-Hermiticity defined in Eq.~(\ref{eta_I}) in both the PT-exact and PT-broken phases. 

\begin{figure}
  \includegraphics[width=\columnwidth]{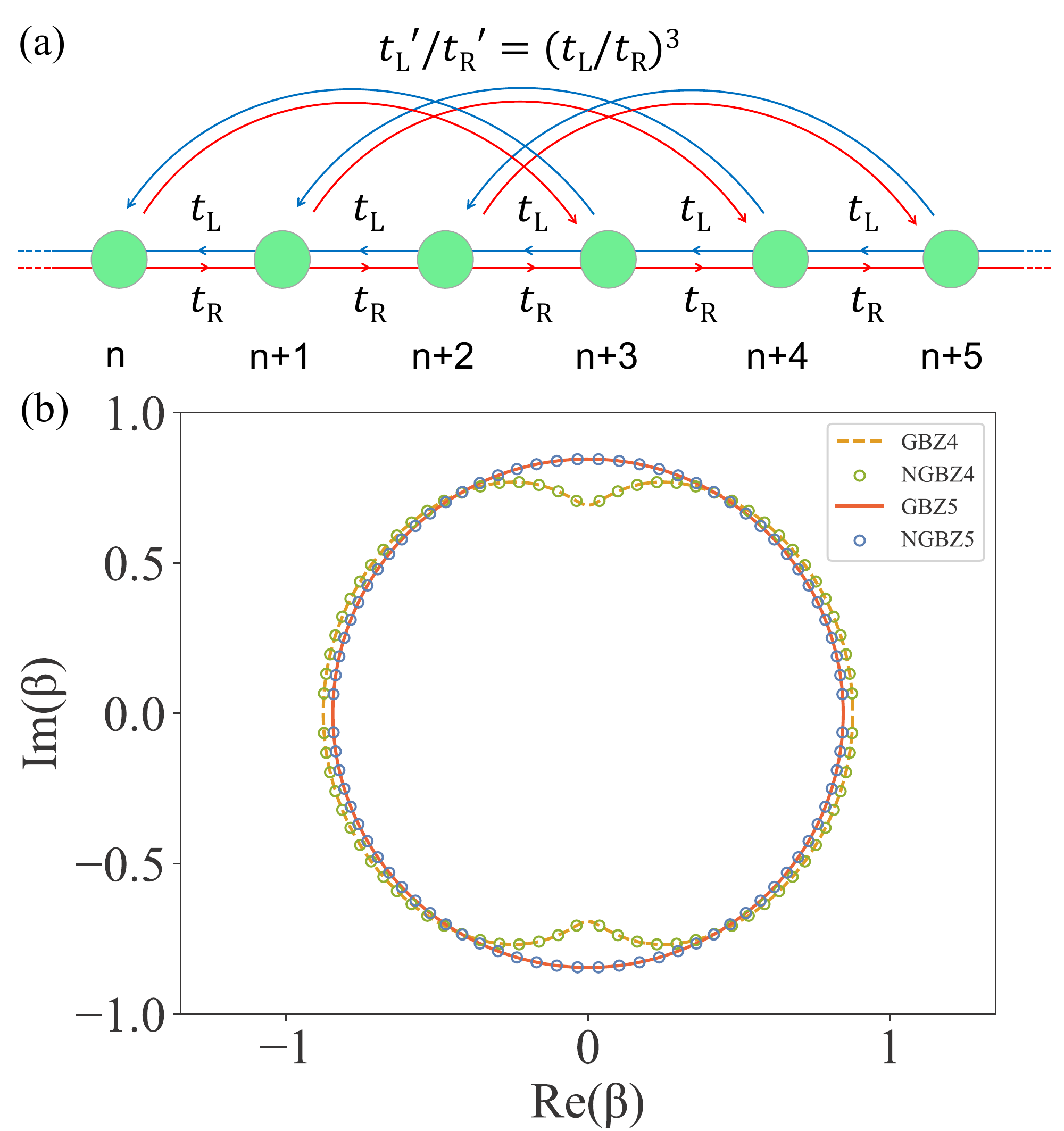}
  \caption{The GBZ of HN model with long-range hopping between the $n$-th and $(n+3)$-th unit cells. (a) The schematic diagram of the model. (b) The GBZ of such a system. The red line represents the GBZ of a system that satisfies the path-independent condition, and the yellow dashed line represents the GBZ of a system that violates the path-independent condition. The parameters are taken as $N=40, t_\mathtt{R} = 0.35, t_\mathtt{L}=0.25$. The long-range hopping strength are taken as $t^\prime_\mathtt{R}=0.1$ and $t^\prime_\mathtt{L}=t^\prime_\mathtt{R} (t_\mathtt{L}/t_\mathtt{R})^3$ to fulfill the path-independent condition. An additional 0.014 is added on $t^\prime_\mathtt{L}$ to display the violated case.}
  \label{fig:3}
\end{figure}

This condition can be interpreted as a path-independence requirement for the product of asymmetric hopping strength ratios between any two sites. In systems with only NN hopping, the path between any two sites is unique, allowing for the application of the IGT. When longer-range hoppings exist, different hopping paths with the same beginning and ending points interfere, and $\eta_{\mathtt{I}}$-pseudo-Hermiticity is only preserved when the nonreciprocity along different paths is the same. A simple example is a HN model with an additional hopping term between $n$th and $(n+3)$th unit cell shown in Fig.~\ref{fig:3}(a), where the path-independence condition implies that the asymmetric ratio $t_\mathtt{R}^\prime / t_\mathtt{L}^\prime$ equals to the product of asymmetric ratios of three NN hoppings $(t_\mathtt{R} / t_\mathtt{L})^3$. Numerical verification of the condition is performed in this example as shown in Fig.~\ref{fig:3}(b). When the asymmetric ratio of long-range hopping equals the product of NN asymmetric ratios, the GBZ is a circle, and the skin modes exhibit the same localization length that is described by the theoretical exponential envelope, a consequence of $\eta_{\mathtt{I}}$-pseudo-Hermiticity. However, when the condition is not satisfied, as shown by the yellow line in Fig.~\ref{fig:3}(b), the GBZ is no longer circular, and therefore, the localization length is no longer unified for different skin modes. This condition also applies in higher dimensions. An example of a 2D square lattice shown in Fig.~\ref{fig:6}(b) will be discussed in detail in Sec.~\ref{sec:2D}. 

Here we provide a physical interpretation of the above condition. When the IGT was first proposed in Ref.~\cite{Hatano_1996}, its physical meaning was explained as an imaginary vector potential. Consider applying an IGT to a Hermitian Hamiltonian to obtain a non-Hermitian one. The imaginary phases introduced by the IGT are the logarithms of asymmetric ratios (see Appendix~\ref{app:hn}). Since the nonreciprocity (the imaginary phase angle) is induced by an imaginary vector potential, the difference in wavefunction before and after the IGT should only differ by an imaginary phase angle dependent on position. This imaginary phase of wavefunction gained between different sites should be path independent as a result of the single-value nature of wavefunction, which leads to our result that the product of asymmetric ratios between different sites should be path-independent.

In addition, even when the system lacks spatial periodicity or when local perturbations disrupt periodicity, a similarity transformation of the diagonal form (which no longer exhibits exponential form because of the broken periodicity) can still be applied using the same procedure, provided that the aforementioned path-independent condition holds.

\section{\label{sec:app}Applications}
\subsection{\label{sec:SSH3}BBC in non-Hermitian SSH3 model}
In the context of non-Hermitian systems, BBC refers to two distinct concepts: the correspondence between the NHSE and energy topology in BZ, the correspondence between conventional topological boundary states and the wave function topology in GBZ \cite{AGBZ_2020}. This section focuses on the latter, as the behavior of the NHSE has already been investigated through IGT. The result that the GBZ is always a circle with radius $|\beta| = |\frac{t_{\mathtt{R}_1}\cdots t_{\mathtt{R}_M}}{t_{\mathtt{L}_1}\cdots t_{\mathtt{L}_{M}}\beta}|$ facilitates the calculation of the topological number in GBZ. We propose a nonreciprocal NN hopping model with three sublattices in one unit cell, referred to as the non-Hermitian trimer SSH (SSH3) model, and determine the topological number corresponding to the number of conventional topological boundary states.

Unlike the well-known SSH model with chiral symmetry, where the topological phase transition occurs at the band touching point and can be well described by the change of Zak's phase \cite{Delplace_2011}, the SSH3 model does not respect chiral symmetry in general and, therefore, cannot be described by topological number defined with Zak's phase. However, discrete energy levels whose eigenstates are boundary states indeed exist \cite{Martinez_2019}, indicating a similar topological origin to topological boundary states in the SSH model. The presence or absence of topological boundary states in the SSH3 model depends on the relative values of the intercell and intracell hopping parameters. When the intercell hopping $t_3$ is less than the intracell hopping $t_1, t_2$, there are no boundary states; when $t_3>t_1, t_3>t_2$, there are two boundary states developed from the middle band and one boundary state each developed from the other two bands; when $t_1 < t_3 < t_2$ or $t_2 < t_3 < t_1$, there is one boundary state each developed from the top and bottom band \cite{SSH3_2022}.

Reference \cite{SSH3_2022} unveils the point chiral symmetry in the SSH3 model and establishes the BBC for the Hermitian SSH3 model with the topological number named normalized sublattice Zak's phase (NS Zak's phase), which is defined as
\begin{equation}
    Z^{\lambda} = -\oint_{\mathtt{BZ}}dk \langle \widetilde{\psi}_\lambda(k)|\partial_k{ \widetilde{\psi}_\lambda (k)} \rangle = -\oint_{\mathtt{BZ}} \frac{\partial{\theta^{\lambda}}}{\partial{k}} dk,
\end{equation}
where $\lambda$ labels the energy band, $|\widetilde{\psi}_\lambda(k)\rangle$ is the projection of wavefunction on the first sublattice divided by the normalization factor (namely, the normalized sublattice wave function), and $\theta^\lambda$ is the relative phase of PBC wavefunction of band $\lambda$ between the first and last sublattices. The number of conventional (Hermitian) boundary states equals the sum of NS Zak's phase over all the bands divided by $2\pi$. 

\begin{figure}
    \includegraphics[width=0.75\columnwidth]{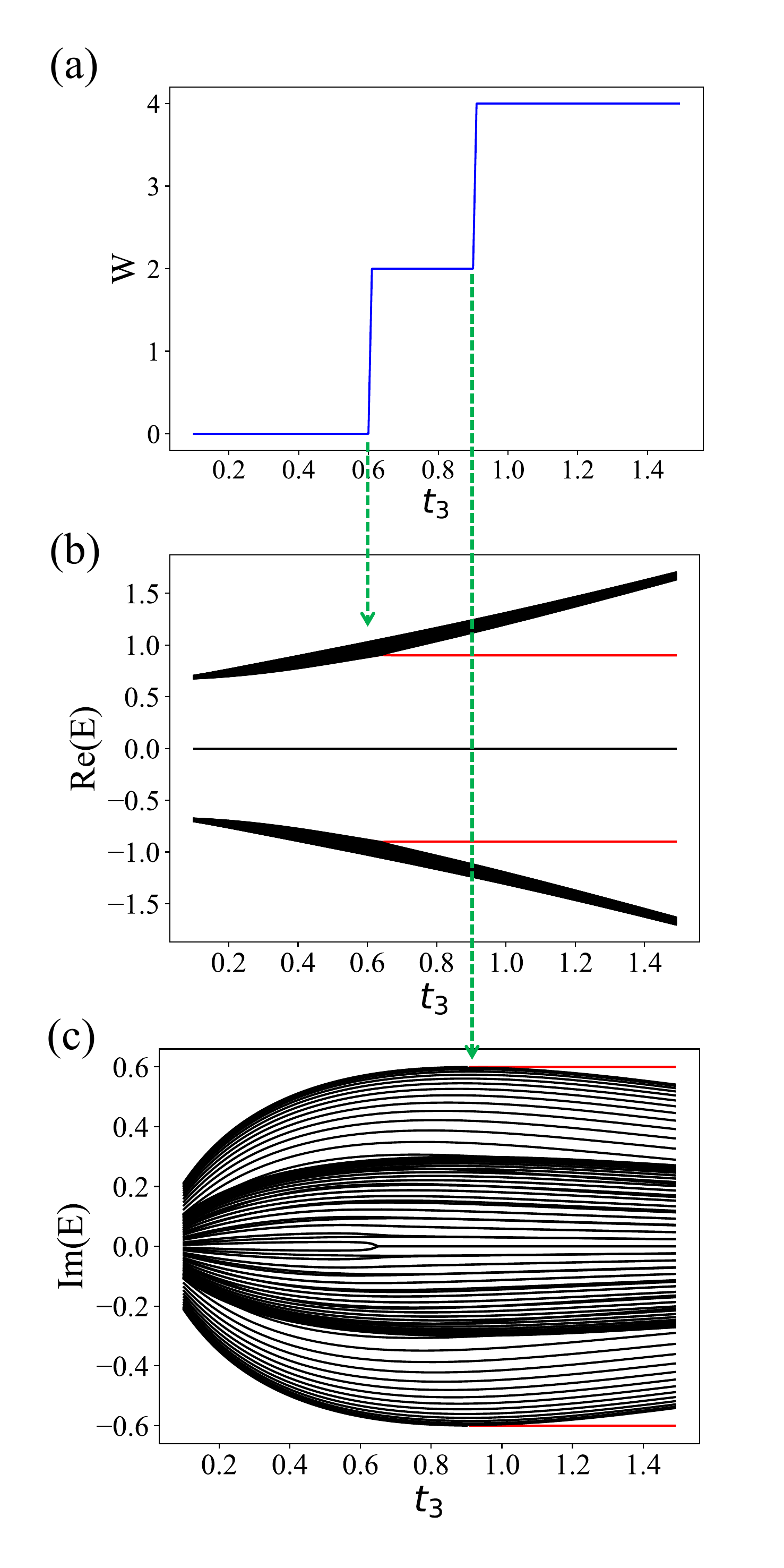}
  \caption{Numerical result of the topological number obtained from the NS Zak's phase and corresponding energy spectrum. The parameters are taken as $N=40, t_\mathtt{L_1}=2.025, t_\mathtt{R_1}=0.4, t_\mathtt{L_2}=-0.4, t_\mathtt{R_2}=0.9$ and $t_\mathtt{L_3}=t_\mathtt{R_3}=t_3$, such that the theoretical transition point locates at $t_3=0.6$ and $t_3 = 0.9$. The real and imaginary parts of the energy spectrum under different $t_3$ are plotted in (b) and (c). The discrete energy levels are marked by the red line. The emergence of discrete levels agrees with the change of topological number displayed in (a).}
  \label{fig:4}
\end{figure}

The non-Hermitian generalization of the SSH3 model corresponds to the $M=3$ case in Sec.~\ref{sec:NHSE}. To determine the number of conventional topological boundary states, the NS Zak's phase is redefined as
\begin{equation} \label{eq:NS}
    \begin{aligned}
       Z^{\lambda}_{\mathtt{nH}} &= -\oint_{\mathtt{GBZ}}dk \langle \widetilde{\phi}_{\lambda}(k)|\partial_k{ \widetilde{\psi}_{\lambda}(k)} \rangle \\
       &= -\frac{1}{2}\oint_{\mathtt{GBZ}}( \frac{\partial{\theta^{\lambda}_R}}{\partial{k}} + \frac{\partial{\theta^{\lambda}_L}}{\partial{k}}) dk \\ 
       &= -\oint_{\mathtt{GBZ}}\frac{\partial {\theta^{\lambda}_R}}{\partial{k}}dk,
    \end{aligned}
\end{equation}
where $|\phi_\lambda(k)\rangle$ denotes the left vector corresponding to $|\psi_\lambda(k)\rangle$, respectively (see Appendix~\ref{app:SSH3} for details). The only difference between the final expression and the Hermitian case lies in calculating the NS Zak's phase within the GBZ.

\begin{figure}
    \includegraphics[width=\columnwidth]{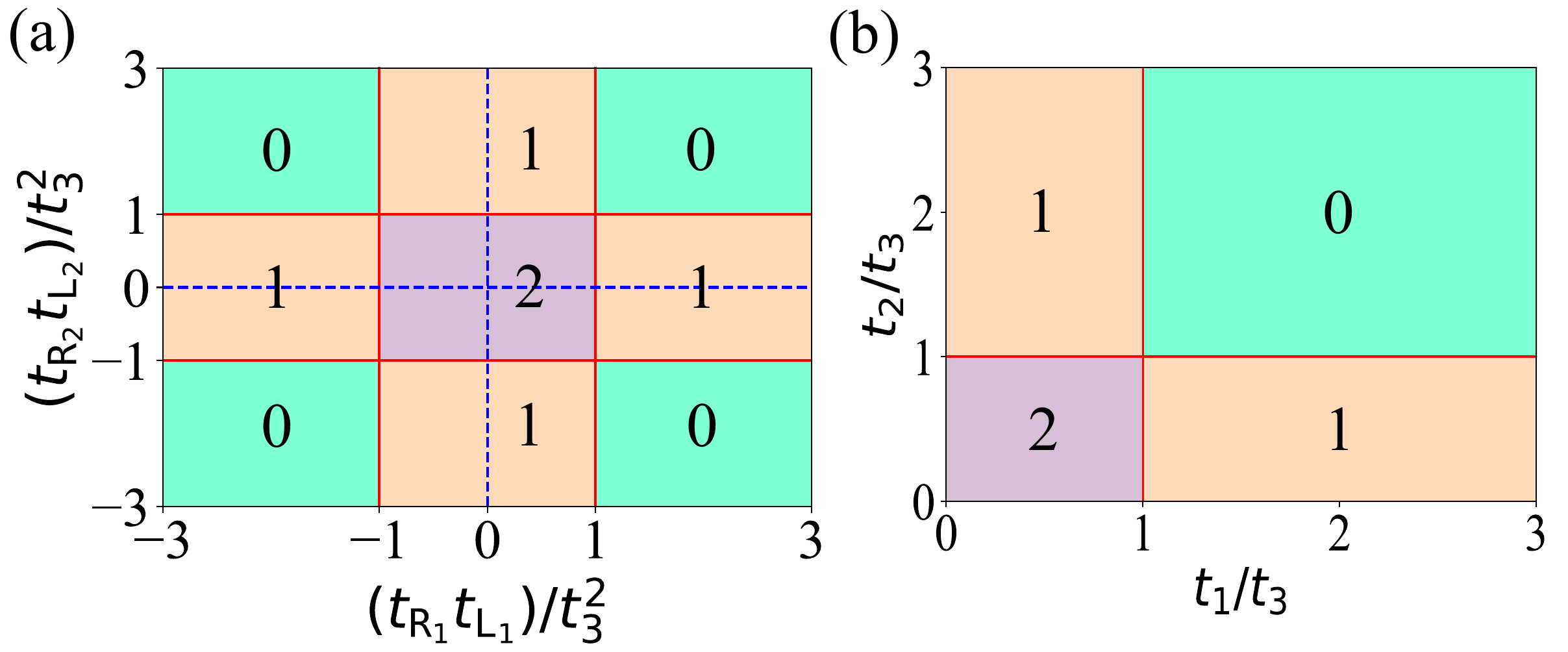}
  \caption{The phase diagrams in both (a) non-Hermitian SSH3 and (b) Hermitian SSH3 model. The red lines display the boundary, and the numbers of edge states at the left edge are labeled in the plot. The blue dashed line in (a) displays the EPs.}
  \label{fig:5}
\end{figure}

We apply Eq.~(\ref{eq:NS}) to calculate the NS Zak's phase and the energy spectrum for the non-Hermitian SSH3 model in the PT-broken phase, as depicted in Fig.~\ref{fig:4}. We define the topological number as the NS Zak's phase divided by $2\pi$, which takes integer values. For simplicity, we assume symmetric intercell hopping with positive values, i.e., $t_{\mathtt{R}_3} = t_{\mathtt{L}_3} = t_3 > 0$. As $t_3$ increases, discrete energy levels appear at the point $t_3 = \sqrt{|t_{\mathtt{L}_1} t_{\mathtt{R}_1}|}$ and $t_3 = \sqrt{|t_{\mathtt{L}_2} t_{\mathtt{R}_2}|}$, which is correctly predicted by the change of NS Zak's phase at these points. The phase diagrams of the non-Hermitian and Hermitian SSH3 models are shown in Fig.~\ref{fig:5}. We take the product $t_\mathtt{L_1}t_\mathtt{R_1}$ and $t_\mathtt{L_2}t_\mathtt{R_2}$ in unit of $t_3^2$ as the axes in the phase diagram of non-Hermitian case since different parameter sets with same product are connected by the IGT (a detailed explanation is provided in Appendix~\ref{app:SSH3}). The main difference between non-Hermitian and Hermitian cases is that there exist EPs depicted by blue dashed lines in Fig.~\ref{fig:5}(a). The PT-exact phase lies in the first quadrant, and the PT-broken phase lies in the remaining three quadrants. The parameters of Fig.~\ref{fig:4} are chosen in the fourth quadrant.

\subsection{\label{sec:2D}Corner-Skin effect in 2D HN model}
A distinguishing feature of 2D nonreciprocal Hamiltonians compared to 1D nonreciprocal Hamiltonians is the absence of Hermitian counterparts in general, even when there are only NN hoppings and the Hamiltonian has an entirely real spectrum. This can be attributed to the condition established in Sec.~\ref{sec:NNN}. Unlike in 1D chains, where the path between any two sites is unique within the NN hopping range, 2D systems allow for multiple paths between arbitrary two sites. Our conclusions derived in 1D chains are valid as long as the product of asymmetric ratio along different paths is unified; otherwise, the $\eta_{\mathtt{I}}$-pseudo-Hermiticity is violated. 

In this section, we use the simplest HN model in a 2D square lattice with OBC as an example to present the numerical results and theoretical predictions of the NHSE. The Hamiltonian of such a system reads
\begin{equation}
\begin{aligned}
    H_\mathtt{HN}^\mathtt{2D} =& \sum_{m=1}^{M-1}\sum_{n=1}^{N} (t_\mathtt{R} a_{m+1, n}^{\dagger}a_{m,n} + t_\mathtt{L} a_{m,n}^{\dagger} a_{m+1,n}) \\
    + &\sum_{m=1}^{M}\sum_{n=1}^{N-1}(t_\mathtt{U} a_{m, n+1}^{\dagger}a_{m,n} + t_\mathtt{D} a_{m,n}^{\dagger}a_{m, n+1}),
\end{aligned}
\end{equation}
where $t_\mathtt{L}, t_\mathtt{R}$ are hoppings along $x$ axis and $t_\mathtt{U}, t_\mathtt{D}$ are hoppings along $y$ axis. The product of asymmetric ratio between two arbitrary sites $(i,j)$ and $(m,n)$ is $(t_\mathtt{U}/t_\mathtt{D})^{m-n}/(t_\mathtt{R}/t_\mathtt{L})^{n-j}$, which is path independent. Hence, we can apply the IGT (in the PT-exact phase)
\begin{equation}
     c_{m,n}^{\dagger} = r_x^m r_y^n a_{m,n}^{\dagger}, \quad c_n = r_x^{-m}r_y^{-n} a_{n},
\end{equation}
where $r_x = \sqrt{t_\mathtt{R}/t_\mathtt{L}}, r_y = \sqrt{t_\mathtt{U}/t_\mathtt{D}}$, to obtain the Hermitian counterpart as well as the energy spectrum. The NHSE can be interpreted as the bulk state modulated by the exponential envelope $r_x^xr_y^y$, which repels all the eigenstates to the corner. The localization lengths along two axes are
\begin{equation} \label{eq:length}
    l_x^{-1} = \frac{1}{2}|\ln|\frac{t_\mathtt{R}}{t_\mathtt{L}}||, \quad  l_y^{-1} = \frac{1}{2}|\ln|\frac{t_\mathtt{U}}{t_\mathtt{D}}||.
\end{equation}
\begin{figure}
	\includegraphics[width=\columnwidth]{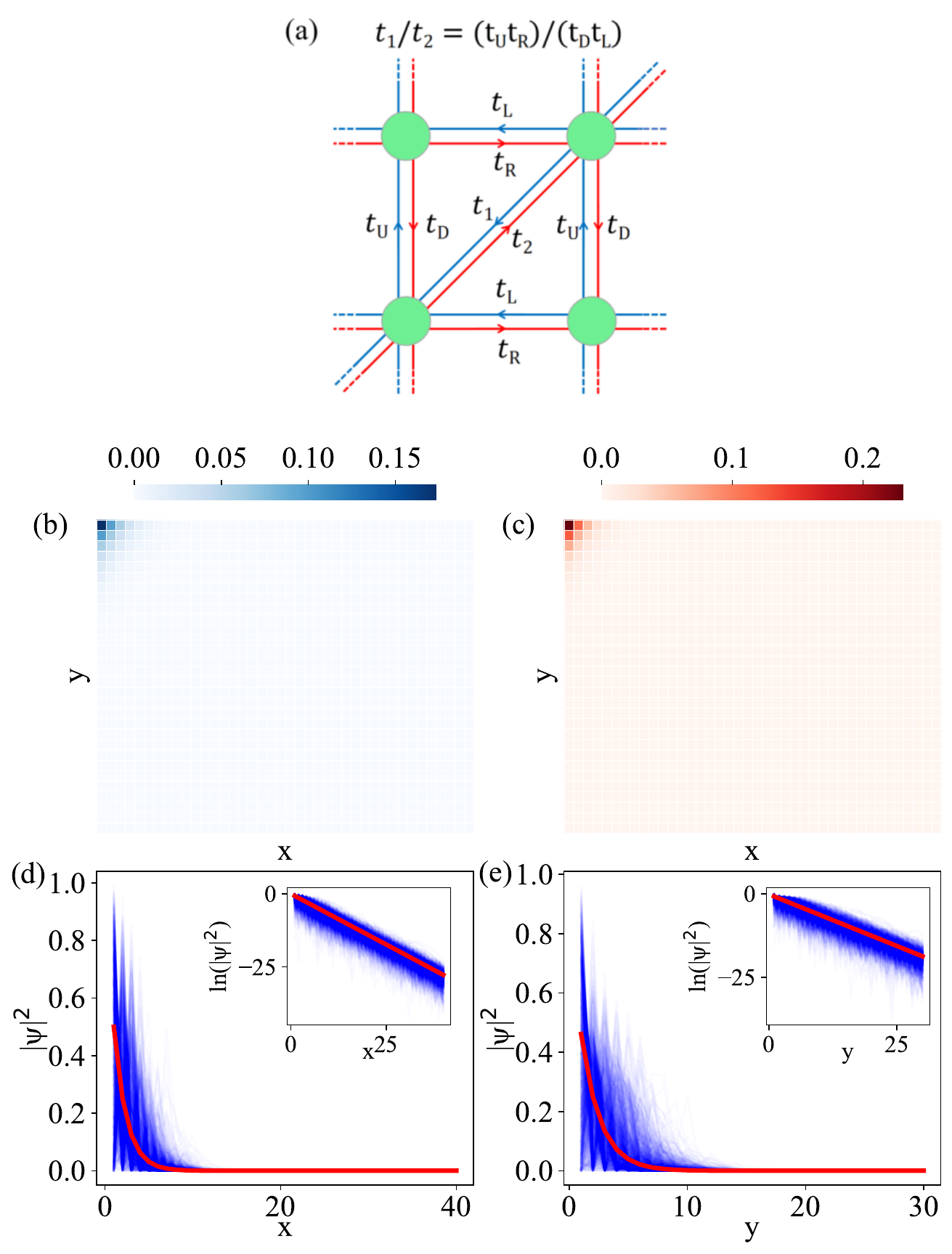}
	\caption{(a) Schematic diagram of 2D HN model with diagonal hopping.  The probability distribution of eigenstates is shown in (b)--(e), and the system size is taken as $N=40, M=30$. (b) The 2D probability distribution of eigenstates. To get all eigenstates in one plot, the plotted distribution is the average of all eigenstates. (c) The theoretical exponential envelope obtained from IGT conforms to the numerical result shown in (b). [(d),(e)] The 1D slice of eigenstates obtained at $y=1$ ($x=1$) is represented by blue lines, and the theoretical exponential envelope is represented by the red line. The inset plot shows the result on a logarithmic scale. Throughout (b)--(e), $t_\mathtt{L}=0.4,t_\mathtt{R}=0.2,t_\mathtt{D}=0.65,t_\mathtt{U}=0.35,t_1=0.5$, and $t_2=t_1(t_\mathtt{D}t_\mathtt{L})/(t_\mathtt{U}t_\mathtt{R})$.}
	\label{fig:6}
\end{figure}
While the energy spectrum is no longer straightforwardly accessible in the PT-broken phase, our theoretical analysis demonstrates that the localization length of skin modes of the PT-broken phase is the same as in the PT-exact phase. Since the generalized Bloch Hamiltonian
\begin{equation}
    H^\mathtt{{2D}}_{\mathtt{HN}}(\beta_x, \beta_y) = t_\mathtt{R}\beta_x^{-1}+t_\mathtt{L}\beta_x + t_\mathtt{U}\beta_y^{-1}+t_\mathtt{D} \beta_y
\end{equation}
has the form of separation of variables. The GBZ of such a system can be constructed analogously to the 1D case, as the characteristic equation $|H^\mathtt{2D}_\mathtt{HN}(\beta_x, \beta_y)-E| = 0$ also exhibits the form of separation of variables \cite{2DGBZ_2023}. Consequently, our findings regarding the unified behavior of skin modes and the identical shape of GBZ in both PT-exact and PT-broken phases, derived from the 1D GBZ theory, remain valid. This facilitates the straightforward acquisition of the continuum bands of the 2D HN model in both phases. 

IGT can be applied to more complicated 2D systems, such as the square lattice with next-nearest-neighbor (NNN) hoppings on the diagonal line shown in Fig.~\ref{fig:6}(a), as long as the path-independent condition of asymmetric ratio is satisfied. In Fig.~\ref{fig:6}(b), we plot the numerical result for the probability distribution of the eigenstates, which agrees well with the theoretical exponential envelope in Fig.~\ref{fig:6}(c). Note that in such a system with NNN hopping, the characteristic equation $|H(\beta_x, \beta_y)-E|=0$ no longer has the form of separation of variables, so we cannot obtain the GBZ directly by decomposing into two 1D systems. However, the IGT approach remains valid even if the GBZ is difficult to obtain. 

\section{\label{sec:discussion} Discussion and outlook}
In this paper, we conduct a systematic study on IGT and the underlying $\eta_{\mathtt{I}}$-pseudo-Hermiticity. By elucidating the unique characteristics of $\eta_{\mathtt{I}}$, we extend the conclusion that all skin modes share the same localization length in the PT-exact phase resulting from IGT to the PT-broken phase. We prove that the GBZ of $\eta_{\mathtt{I}}$-pseudo-Hermitian Hamiltonian is a perfect circle, which paves the way to obtain the continuum band properties and investigate the wave function topology in GBZ. This conclusion is still valid when the hopping strength takes a complex value. We apply our result to the non-Hermitian SSH3 model and obtain the non-Hermitian generalization of NS Zak's phase to establish BBC and obtain the whole phase diagram. We further generalize the condition of $\eta_{\mathtt{I}}$-pseudo-Hermiticity from NN hopping to the path-independence of the product of asymmetric ratio. We exemplify the 2D HN model to present the application of IGT in 2D systems that satisfy such conditions. 
\subparagraph{}
Here we highlight several potential directions for future work, some relevant experimental realizations, and observable effects. The procedure of establishing the BBC for the non-Hermitian SSH3 model based on the generalization of NS Zak's phase can be extended to a more general SSHm model. A rigorous theoretical proof for such a generalization is also needed. Another topic is applying IGT to 2D systems, which satisfy the path-independent condition. Since the GBZ theory for general 2D systems is still unclear, the straightforward approach of similarity transformation can help investigate the energy spectra, boundary states, and GBZs when the 2D systems cannot be regarded as two separable 1D systems.

The nonreciprocal lattice has been realized in mechanical \cite{Brandenbourger_2019non, Gao_2020anomalous, Ghatak_2020observation, Zhang_2021acoustic, Gao_2022non, Wang_2023extended}, electrical \cite{Helbig_2020, Hofmann_2020, Zou_2021}, and optical \cite{Weidemann_2020light, Song_2020twodimensional, Wang_2021braid, Kai_2021winding, Weidemann_2022topological} systems. By tuning the nonreciprocal parameter, our result of unified localization lengths in both phases can be verified. Since these systems are generally extendable to models with more sublattices, we expect experimental investigation of the NHSE in generic nonreciprocal models with multiple sublattices and the BBC in the nonreciprocal SSH3 model established in this paper. Besides classical realizations, the analog quantum simulation of non-Hermitian models has been realized in ultracold atomic platforms to investigate the interplay between NHSE and many-body physics \cite{Li_2020topological, Liang_2022dynamic, Lin_2023}. Although the behavior of systems investigated in this paper is simple as a result of circular GBZ, it would be interesting to investigate the behaviors with the existence of interactions. Furthermore, the digital quantum simulation of non-Hermitian systems is also a topic of interest. The main difficulty in simulating a non-Hermitian system is the implementation of non-unitary evolution, which is usually realized by post-selection \cite{Long_2006, Long_2008, Long_2009, Cui_2012, Wei_2016, Shao_2019, Wen_2019, Zheng_2021}. The digital quantum simulation may experimentally demonstrate our results in models beyond NN hopping by specifically designing nonreciprocity to satisfy the path-independent condition.

We also note recent paper on the Aharonov-Bohm effect for imaginary magnetic fields \cite{Tomoki_2024}, in which the imaginary vector potential amplifies or decays the amplitude of the wavefunction after winding a close loop. For generic lattice Hamiltonian amenable to the IGT, the physical explanation of such IGT is adding an imaginary vector potential. Thus, we expect to observe this effect in the 2D HN model with diagonal hoppings.

\begin{acknowledgments}
The authors would like to thank Xiaogang Li, Yuanye Zhu, and Pengyu Wen for their helpful discussion. This work is supported by the National Natural Science Foundation of China under Grants No. 11974205, and No. 61727801, and the Key Research and Development Program of Guangdong Province (Grant No. 2018B030325002).
\end{acknowledgments}

\appendix

\section{\label{app:hn} Applying IGT to HN models with real spectra}
The IGT was first applied in the HN model \cite{Hatano_1996}, which can be expressed as 
\begin{equation}
    H_\mathtt{HN} = \sum_{n=1}^{N-1}(t_\mathtt{R} a_{n+1}^{\dagger}a_{n} + t_\mathtt{L} a_{n}^{\dagger} a_{n+1}),
\end{equation}
where $a_{n}^{\dagger}$ ($a_{n}$) are the creation (annihilation) operators at site $n$, the parameters $t_\mathtt{R},t_\mathtt{L}\in\mathbb{R}$ are the asymmetric hopping amplitudes and $N$ is the system size. It can be transformed to a Hermitian Hamiltonian $H^{\prime}$ with reciprocal hopping term $t^{\prime} = \sqrt{t_\mathtt{L}t_\mathtt{R}} $ when $t_\mathtt{R}t_\mathtt{L}>0$, by taking an imaginary phase angle $\theta = i\ln(\sqrt{{t_\mathtt{R}}/{t_\mathtt{L}}})$ in the following gauge transformation \cite{Hatano_1996, Yao_2018},
\begin{equation}
    c_n^{\dagger} = e^{-in\theta} a_{n}^{\dagger}, \quad c_n = e^{in\theta} a_{n}.
\end{equation}
Or alternatively, we have $H^{\prime}=S^{-1} H_\mathtt{NH} S$, where $S$ is a diagonal matrix whose diagonal elements are $\{r, r^2, \cdots, r^N\}$ with $r = \sqrt{{t_\mathtt{R}}/{t_\mathtt{L}}}$. The non-Hermitian Hamiltonian $ H_\mathtt{NH}$ has an entirely real spectrum as this similarity transformation does not change the spectrum. Since the IGT is no longer unitary, it acts as a rescaling of the eigenstates, which leads to the NHSE. To be more concrete, a bulk eigenstate $\left\vert \bar{\psi_{l}}\right\rangle$ of Hermitian $ H^{\prime}$ is an extended Bloch wave due to the discrete translation symmetry of the bulk sites; therefore, $H_\mathtt{NH}$'s eigenstate $\left\vert \psi_{l}\right\rangle=S\left\vert \bar{\psi_{l}}\right\rangle$ is exponentially
localized at the left (right) edge of the chain for $r<1$ $(r>1)$  with localization length $\left\vert \ln r\right\vert ^{-1}$. The above IGT is limited at the parameter region $t_\mathtt{L} t_\mathtt{R} > 0$, which corresponds to the PT-exact phase.

\section{\label{app:pseudo}Important conclusions on $\eta$-pseudo-Hermitian Hamiltonians}
In this section, we give a brief explanation of the conclusions on $\eta$-pseudo-Hermitian Hamiltonians used in Sec.~\ref{sec:pseudo}. The detailed derivation of these results can be found in Ref.~\cite{Mostafazadeh_2002pseudo}. The Hermitian operator $\eta$ can define a new inner product as
\begin{equation}
    \langle \phi | \psi \rangle_{\eta} = \langle \phi | \eta | \psi \rangle,
\end{equation}
where $| \psi \rangle,| \phi \rangle$ are arbitrary vectors in the Hilbert space. The $\eta$-pseudo-Hermitian Hamiltonians are equal to their adjoints under this new inner product. When $\eta$ is the identity operator, Eq.~(\ref{eq:eta_def}) reduces to the standard definition of Hermiticity. Therefore, $\eta$-pseudo-Hermiticity can be regarded as a generalization of Hermiticity.

The $\eta$-pseudo-Hermiticity means $H = \eta^{-1} H^{\dagger} \eta$, hence the eigenstates of $H$ and $H^{\dagger}$ with same eigenvalues are connected by the $\eta$ transform. To be more explicit, $\eta|\psi_i\rangle$ is the eigenstate of $H^\dagger$ with energy $E_i$ if $|\psi_i\rangle$ is the eigenstate of $H$ with the same energy. Note that for every $|\psi_i \rangle$ with energy $E_i$, there exists a corresponding $| \phi_i \rangle$ with energy $E_i^*$. Therefore, we conclude that the eigenvalues come in either real values or complex conjugate pairs if and only if the Hamiltonian is $\eta$-pseudo-Hermitian. Such spectrum property differs from the Hermitian case, where all eigenvalues are real. The difference can be understood because the inner product defined by $\eta$ can be indefinite.

Another key conclusion in Ref.~\cite{Mostafazadeh_2002pseudo} is the relationship between $\eta$-pseudo-Hermiticity and the widely utilized concept of PT-symmetry \cite{Bender_98PT}. A Hamiltonian respects PT-symmetry if it is invariant under the PT transformation, where P is the parity operator, and T is the time-reversal operator, respectively. A PT-symmetric system is classified as belonging to the PT-exact phase if all its eigenvalues are real or the PT-broken phase if at least one complex conjugate pair exists. Since the eigenvalues of PT-symmetric Hamiltonians are either real or complex-conjugate pairs, PT-symmetric Hamiltonians are all $\eta$-pseudo-Hermitian.

\section{\label{app:reflection-symmetry}Symmetry generated by the $\eta$-pseudo-Hermiticity}
If there exists two distinct $\eta$ operators in a system, $\eta_1 H = H^{\dagger}\eta_1$ and $\eta_2 H = H^{\dagger}\eta_2$, then it can be straightforwardly shown that $[\eta_1^{-1}\eta_2, H]=0$, which means $\eta_1^{-1}\eta_2$ is a symmetry of the systems \cite{Mostafazadeh_2002pseudo}. This property can be utilized to uncover hidden symmetries in non-Hermitian systems. As an illustration, consider the SSH model with reciprocal hopping:
\begin{equation}
    H_{\mathtt{SSH}} = \sum_{n=1}^{N}  t_1 b_n^{\dagger
} a_n +\sum_{n=1}^{N-1}t_2 a_{n+1}^{\dagger} b_n + H.c.,
\end{equation}
where $a_n^{\dagger}$ ($a_n$) and $b_n^{\dagger}$ ($b_n$) are the creation (annihilation) operators for the two sublattices in the $n$-th unit cell. Such a model respects mirror reflection symmetry in real space, as described by the matrix:
\begin{equation}
\widetilde{\mathcal{R}} = \begin{pmatrix}
            & & & 1 \\
            & & \begin{turn}{80}$\ddots$\end{turn} & \\
            & 1 & & \\
            1 & & & \\
        \end{pmatrix}.
\end{equation}
The non-Hermitian SSH model with nonreciprocal hopping terms, corresponding to the $M=2$ case of our general nonreciprocal Hamiltonian, breaks the above mirror reflection symmetry. More precisely, the reflection operation $\widetilde{\mathcal{R}}$ applied to the general model results in the transformation $t_{\mathtt{L}_{i}} \leftrightarrow t_{\mathtt{R}_{M-i}}$ for $i=1,\cdots,M-1$ and $t_{\mathtt{L}_M}\leftrightarrow t_{\mathtt{R}_M}$ . The Hermitian SSH Hamiltonian, which is the $M=2$ and $t_\mathtt{L}=t_\mathtt{R}$ case, remains invariant under this transformation, with $t_1 \leftrightarrow t_1$ and $t_2 \leftrightarrow t_2$. However, the nonreciprocal SSH Hamiltonian is transformed to its Hermitian conjugate under this transformation, with $t_{\mathtt{L}_1}\leftrightarrow t_{\mathtt{R}_1}$ and $t_{\mathtt{L}_2}\leftrightarrow t_{\mathtt{R}_2}$. Consequently, $\widetilde{\mathcal{R}}$ no longer acts as a symmetry operator but instead becomes an $\eta$ operator. Notably, the eigenvalues of $\widetilde{\mathcal{R}}$ are $\pm 1$, rendering it indefinite. This exemplifies the existence of indefinite $\eta$  for the real spectrum. Since we now have two $\eta$ operators with different physical meanings, one $\eta_{\mathtt{I}}$ that describes the exponential rescaling of inner-product space and the other $\widetilde{\mathcal{R}}$ that describes the reflection along the middle point, a new symmetry can be generated as
\begin{equation}
    g_{\mathtt{SSH}} = \widetilde{\mathcal{R}}_N \cdot \mathtt{diag}\{r_2^{-2}, r_2^{-4}, \cdots, r_2^{-2N}\} \otimes
        \begin{pmatrix}
            0 & r_{1}^{-2} \\
            r_0^{-2} & 0 \\
        \end{pmatrix},
\end{equation}
where $\mathcal{R}_N$ represents the $N$ by $N$ reflection matrix. This novel symmetry for the nonreciprocal SSH model is intimately linked to the broken mirror reflection symmetry induced by the nonreciprocal term, facilitated by $\eta_{\mathtt{I}}$ generated through the IGT. 

The presented procedure can be extended to encompass general models with multiple sublattices. By decomposing the hopping terms into reciprocal and nonreciprocal components,  we can express them as $t_{\mathtt{L}_{i}}=t_i-\gamma_i, t_{\mathtt{R}_{i}}=t_i+\gamma_i$, respectively. If the reciprocal component exhibits reflection symmetry, i.e., $t_i = t_{M-i}$, and the nonreciprocal part component adheres to the same constraint, $\gamma_i = \gamma_{M-i}$, then the reflection operator $\widetilde{\mathcal{R}}$ fulfills the definition of $\eta$, and therefore yields the new symmetry $g$ as
\begin{equation}
    \begin{aligned}
        g = \widetilde{\mathcal{R}}_{MN} \eta_{\mathtt{I}}.
    \end{aligned}
\end{equation}

\section{\label{app:eta}Details on the general expression of $\eta$}
In this section, we provide more details on the expression of the $\eta$ operator given in Eq.~(\ref{eq:eta_F}). Note that it is different from Eq.~(22) in Ref.~\cite{Mostafazadeh_2002pseudo} since $c_{i_0}$ can take $\pm 1$. The HN model is first examined in both PT-exact and PT-broken phases to illustrate the applicability of Eq.~(\ref{eq:eta_F}). The OBC eigenstate for the HN model takes the form
\begin{equation}
    |\psi\rangle = \sum_{n=1}^N (a\beta_1^n+ b\beta_2^n) |n\rangle,
\end{equation}
where $\beta_1, \beta_2$ are two points in the GBZ corresponding to the same energy, and $a,b$ denote the superposition coefficients determined by the boundary condition $\langle 0|\psi\rangle = \langle N+1 | \psi\rangle=0$. Since the GBZ is a circle with radius $r =\sqrt{|t_\mathtt{R}/t_\mathtt{L}|}$, it can be parameterized as $\beta(k) = re^{ik}$ with $k \in [0, 2\pi)$. To obtain $\beta_1$ and $\beta_2$, the model needs to be analyzed separately in the PT-exact and PT-broken phases. The system is in the PT-exact phase when $ \omega \equiv t_\mathtt{L} t_\mathtt{R} > 0$ and in the PT-broken phase when $\omega <0 $. The energy spectrum is either real or imaginary, which has the form
\begin{equation}\label{eq:energy}
    E(k) = \begin{cases}
                2r \cos k \quad \omega > 0,\\
                2ir   \sin k \cdot \mathtt{sgn} (t_L) \quad \omega < 0.
            \end{cases}
\end{equation}
Hence, the parameters $k_1$ and $k_2$  corresponding to the same energy in the PT-exact phase fulfill  $k_1 = -k_2$, while in the PT-broken phase, they satisfy $k_1 + k_2 = \pi$. The expression of the eigenstates corresponding to Eq.~(\ref{eq:energy}) is given by 
\begin{equation}
    |\psi(k)\rangle = \begin{cases}
                            \sqrt{\frac{2}{N+1}}\sum_{n=1}^N  r^{n}\mathtt{sin}(nk)|n\rangle \quad \omega > 0, \\
                            \sqrt{\frac{1}{\lfloor N / 2 \rfloor}}\sum_{n=1}^N  r^n (e^{ink} - e^{-in(k+\pi)})|n\rangle\\
                            \omega < 0,
                      \end{cases}
\end{equation}
where $\lfloor N / 2 \rfloor$ is defined as the greatest integer less than or equal to $N / 2$. The possible values for $k$ are determined by the boundary condition, which reads
\begin{equation}\label{eq:k}
    k = \begin{cases}
            \frac{j\pi}{N+1}\quad \omega > 0,\\
            \frac{j\pi}{N+1} + \frac{\pi}{2} \quad \omega < 0,
        \end{cases} \quad j=1,2,\cdots,N.
\end{equation}
Similarly, the eigenstates of $H_\mathtt{NH}^{\dagger}$ can be expressed as
\begin{equation}\label{eq:phi}
    |\phi(k)\rangle = \begin{cases}
                            \sqrt{\frac{2}{N+1}}\sum_{n=1}^N r^{-n}\mathtt{sin}(nk)|n\rangle\quad \omega > 0, \\
                            \sqrt{\frac{1}{\lfloor N / 2 \rfloor}}\sum_{n=1}^N r^{-n} (e^{ink} - e^{-in(k+\pi)})|n\rangle\\
                            \omega < 0.
                      \end{cases}
\end{equation}
We can verify the energies corresponding to the eigenstates $|\psi(k)\rangle$ and $|\phi(k)\rangle$ are complex conjugates. This implies the bi-orthonormal condition $\langle \phi (k) | \psi(k^\prime)\rangle = \delta_{kk^\prime}$, which can also be confirmed by directly calculating their inner product. Note that the normalization coefficients are not unique, i.e., we can take $|\psi(k)\rangle \rightarrow a(k)|\psi(k)\rangle$ and $|\phi(k)\rangle \rightarrow |\phi(k)\rangle/a(k)$ while the bi-orthonormal condition is still fulfilled. Therefore, Eq.~(\ref{eq:eta_F}) can lead to different metrics. Now we substitute $|\phi(k)\rangle$ expressed as Eq.~(\ref{eq:phi}) into Eq.~(\ref{eq:eta_F}) to confirm that it indeed produces $\eta_{\mathtt{I}}$. In the PT-exact phase, the result is given by
\begin{equation}\label{eq:A6}
    \begin{aligned}
        \eta &= \sum_{k} |\phi(k)\rangle \langle \phi(k)| \\
        &= \frac{2}{N+1} \sum_{m,n,k} r^{-(m+n)} \mathtt{sin}(mk)\mathtt{sin}(nk)|m\rangle \langle n|.
    \end{aligned}
\end{equation}
Only terms with $m=n$ in Eq.~(\ref{eq:A6}) take nonzero values. This can be easily understood in the thermodynamic limit, where the summations over $k$ are replaced by integrals. Then, the expression can be simplified as 
\begin{equation}
    \begin{aligned}
        \eta &= \frac{2}{N+1} \sum_{n,k} r^{-2n} \mathtt{sin}^2(nk)|n\rangle \langle n| \\
        &= \sum_{n} r^{-2n} |n\rangle \langle n|,
    \end{aligned}
\end{equation}
which equals to $\eta_{\mathtt{I}}$. The above derivation shows that $\eta_{\mathtt{I}}$ satisfies Eq.~(\ref{eq:positive}) in PT-exact phase, which we theoretically proved in the main text by the fact $\eta_{\mathtt{I}}$ is positive definite.

In the PT-broken phase, we need to choose the eigenstates $|\phi(k)\rangle$ and $|\phi(k^\prime)\rangle$ with conjugate energies. From Eq.~(\ref{eq:energy}) and Eq.~(\ref{eq:k}), we have $E(k) = E(k^\prime)^*$ for $k+k^\prime = 2\pi$. Therefore, $\eta$ can be expressed as 
\begin{widetext}
\begin{equation}
    \begin{aligned}
        \eta =\frac{1}{\lfloor N/2 \rfloor} \sum_k& (|\phi(k)\rangle \langle \phi(2\pi-k)| + |\phi(2\pi-k)\rangle \langle \phi(k)| )\\ 
        = \frac{1}{\lfloor N / 2 \rfloor}\sum_{j,m,n}&[ r^{-2(m+n)}(r^2\mathtt{cos}(2n-1)\theta_j \ \mathtt{cos}(2m-1)\theta_j |2m-1\rangle\langle 2n-1|-\mathtt{sin}2n\theta_j \ \mathtt{sin}2m\theta_j|2m\rangle\langle 2n|)   \\ 
        &+ir^{-2(m+n)+1}(\mathtt{cos}(2n-1)\theta_j \ \mathtt{sin}2m\theta_j|2m\rangle\langle 2n-1|+\mathtt{cos}(2m-1)\theta_j \ \mathtt{sin}2n\theta_j|2m-1\rangle\langle 2n|)] ,
    \end{aligned}
\end{equation}
\end{widetext}
where $\theta_j = j\pi/(N+1)$. Similarly to Eq.~(\ref{eq:A6}), the summation in every off-diagonal term equals zero. The remaining diagonal terms yield
\begin{widetext}
\begin{equation}
\begin{aligned}
    \eta &= \frac{1}{\lfloor N / 2 \rfloor}\sum_{j,n} r^{-2(2n-1)}\mathtt{cos}^2(2n-1)\theta_j\ |2n-1\rangle\langle 2n-1| - r^{-4n} \mathtt{sin}^2 2n\theta_j |2n\rangle\langle 2n| \\ 
    &= \sum_n r^{-2n} |n\rangle\langle n|,  
\end{aligned}
\end{equation}
\end{widetext}
which also equals to $\eta_\mathtt{I}$. Since the spectrum is entirely imaginary for the HN model in the PT-broken phase, we do not need to consider the values of $c_{i_0}$. For a general lattice model, the spectrum in the PT-broken phase is not always entirely imaginary, hence the values of $c_{i_0}$ need to be considered. In the subsequent analysis, we show that any $\eta$ operator can be expressed in the form of Eq.~(\ref{eq:eta_F})
by appropriate selection of $c_{i_0}$.

After verifying that Eq.~(\ref{eq:eta_F}) leads to $\eta_{\mathtt{I}}$ in the HN model,  we give a theoretical explanation of how an arbitrary $\eta$ operator can be expressed by Eq.~(\ref{eq:eta_F}). Equation~(\ref{eq:eta_F}) is obtained by substituting the 
relations $\eta|\psi_{i_{\pm}}\rangle=c_{i_{\pm}}|\phi_{i_{\mp}}\rangle$ and $\eta|\psi_{i_{0}}\rangle=c_{i_{0}}|\phi_{i_{0}}\rangle$ into
completeness of bi-orthonormal basis
\begin{equation}\label{eq:completeness}
    \sum_{i_{\pm}} |\psi_{i_{\pm}}\rangle \langle \phi_{i_{\pm}}|+ \sum_{i_{0}} |\psi_{i_{0}}\rangle \langle \phi_{i_{0}}| = 1
\end{equation}
 and taking $c_{i_\pm}=1$. To achieve this for an arbitrary bi-orthonormal basis, we need to adjust the normalization coefficients. After attempting to normalize all $c_{i_{\pm}}$ and $c_{i_{0}}$ to unity, we demonstrate that only $c_{i_\pm}=1$ can be consistently set to $1$, while $c_{i_0}$
  can only take values of $1$ or $-1$. To maintain the bi-orthonormal condition, the transformation of the eigenstates should have the form of
\begin{equation}\label{A11}
    |\tilde{\psi}_{i_{\pm,0}}\rangle= a_{i_{\pm,0}}|\psi_{i_{\pm,0}}\rangle, \quad |\tilde{\phi}_{i_{\pm,0}}\rangle= |\phi_{i_{\pm,0}}\rangle/a_{i_{\pm,0}}^*,
\end{equation}
where $a_{i_{\pm}}$ and $a_{i_{0}}$ denote the adjustment on normalization coefficients. Then we substitute Eq.~(\ref{A11}) into Eq.~(\ref{eq:coeff}) and try to obtain
\begin{equation}
    \eta|\tilde{\psi}_{i_{\pm}}\rangle = |\tilde{\phi}_{i_{\mp}}\rangle, \quad \eta|\tilde{\psi}_{i_{0}}\rangle = |\tilde{\phi}_{i_{0}}\rangle
\end{equation}
which straightforwardly leads to
\begin{equation}
   c_{i_\pm} = \frac{1}{a_{i_\pm}a_{i_\mp}^*},\quad c_{i_0} = \frac{1}{|a_{i_0}|^2}.
\end{equation}
The former relation requires that $c_{i_\pm} = c^*_{i_\mp}$ and the latter relation requires that $c_{i_0} \in \mathbb{R}^+$. On the other hand, by substituting the completeness of bi-orthonormal basis Eq.~(\ref{eq:completeness}) into Eq.~(\ref{eq:coeff}), we obtain another expression for $c_{i_{\pm, 0}}$ and  $c_{i_{0}}$, given by 
\begin{equation}
    c_{i_{\pm}} = \langle\psi_{i_{\mp}}|\eta|\psi_{i_{\pm}}\rangle, \quad c_{i_{0}} = \langle\psi_{i_{0}}|\eta|\psi_{i_{0}}\rangle,
\end{equation}
Due to the Hermiticity of $\eta$, we have 
\begin{equation}
    c_{i_\pm} = c_{i_\mp}^*, \quad c_{i_0} \in \mathbb{R}.
\end{equation}
The condition $c_{i_\pm} = c^*_{i_\mp}$ is always satisfied, implying all coefficients $c_{i_\pm}$ can be set to 1 through appropriate normalization.  While we can make positive $c_{i_0}$ to $1$ and  negative $c_{i_0}$ to $-1$, achieving absolute uniformity (namely all $c_{i_0}$ equal to $1$ ) remains impossible. This necessitates retaining an undetermined coefficient $c_{i_0}$ in Eq.~(\ref{eq:eta_F}). If all $c_{i_0} = 1$, we can straightforwardly show that $\eta$ operator expressed by Eq.~(\ref{eq:eta_F}) is positive definite when the spectrum is entirely real and indefinite when the spectrum is complex (and vice versa). Consider arbitrary nonzero right vector $|f\rangle$, which can be written as
\begin{equation}
    |f\rangle = \sum_{i_{\pm}}|\psi_{i_\pm}\rangle\langle\phi_{i_\pm}|f\rangle +\sum_{i_{0}}|\psi_{i_0}\rangle\langle\phi_{i_0}|f\rangle
\end{equation}
with the help of completeness.  By substituting it into the bi-orthonormal condition, the inner product $\langle f|\eta| f\rangle$ takes the form
\begin{equation}
    \langle f|\eta|f\rangle = \sum_{i_\pm}\langle f|\phi_{i_+}\rangle\langle \phi_{i_-}|f\rangle + \sum_{i_0}|\langle \phi_{i_0}|f\rangle|^2.
\end{equation}
If we have an entirely real spectrum, only the second term exists, and the inner product is, therefore, positive for any nonzero $|f\rangle$. But if we have complex eigenvalues, the first term can give negative values or zeros. For example, we can take $|f\rangle = |\psi_{i^\prime_+}\rangle-|\psi_{i^\prime_-}\rangle$ for arbitrary index $i^\prime$, then the inner product equals --2 and thus $\eta$ is indefinite.
If taking some $c_{i_0}=-1$, we can have indefinite $\eta$ even if the spectrum is entirely real.

\section{\label{app:GBZ}Details on the non-Bloch theory and GBZ}
In this section, we give a brief review of the non-Bloch theory and GBZ introduced by Ref.~\cite{GBZ_2019}. Since the spectra and eigenstates of OBC and PBC Hamiltonians can differ significantly from each other in non-Hermitian systems, the Bloch wave picture is no longer valid in OBC. Nevertheless, extending the wave vector k to a complex value can still solve the system \cite{Yao_2018}. The eigenstates can be expressed as a linear combination of generalized Bloch wave functions
\begin{equation}
    |\psi(E) \rangle = \sum_j \sum_{n=1}^N \beta_j(E)^n |n\rangle \otimes |\mu_j\rangle,
\end{equation}
where $\beta_j=e^{ik_j}$ can go beyond the unit circle when $k_j$ is complex-valued, $|\mu_j\rangle$ is the distribution in different sublattices of one unit cell, and $|n\rangle$ is the lattice basis. Subscript $j$ denotes different $\beta$ with the same energy $E$. The eigenstates of a system are constructed by linearly combining these generalized Bloch wave functions to satisfy the boundary condition. Analogous to the bulk Hamiltonians $H(k)$ in Hermitian cases, $H(\beta)_{M\times M}$ can be given by choosing the generalized Bloch wave function as the basis. Specifically, the generalized Bloch Hamiltonian $H(\beta)$ for the tight-binding model described by Eq.~(\ref{eq:tb}) takes the form of Eq.~(\ref{eq:beta}). Then, the eigenvalue problem in the original real-space Hamiltonian is converted to solving the eigenvalue $E$ and eigenstates $|\mu\rangle$ of $H(\beta)$ if $\beta$ is given. In contrast to Hermitian systems, where the permissible $\beta$ values are confined to the unit circle, non-Hermitian systems exhibit a broader range of $\beta$ values that trace complicated curves on the complex plane, forming the so-called GBZ. The non-Bloch theory points out that the GBZ is determined by the $\beta$ that constructs continnum bands. By substituting the general form of OBC for the 1D non-Hermitian systems without symmetry into the characteristic equation, the GBZ is obtained by the condition $|\beta_p|=|\beta_{p+1}|$, where $\beta_{p}$ and $\beta_{p+1}$ represent the $p$th and $(p+1)$th $\beta$ values respectively, when sorted in ascending order of magnitude by $|\beta_1| \leq |\beta_2| \leq \cdots \leq |\beta_{2p}|$. The GBZ is formed by tracing the trajectory of these $\beta_{p}$ and $\beta_{p+1}$ values across different energy levels within the continuum bands.

\section{\label{app:SSH3} NS Zak's phase in non-Hermitian SSH3 model}
In this section, we provide more details on how to generalize the NS Zak's phase in the non-Hermitian SSH3 model. Predicting the number of conventional edge states (in contrast to skin modes) generally requires two modifications to the expressions of topological number: one is to replace all $\langle \psi|$ with the left vector $\langle \phi|$, where $|\psi\rangle$ and $|\phi\rangle$ are eigenstates of $H$ and $H^{\dagger}$ with conjugate eigenvalues; the other is that the calculation should be made in GBZ instead of BZ \cite{Yao_2018}. In the case of NS Zak's phase in the SSH3 model, the original expression valid for the Hermitian case is 
\begin{equation}\label{eq:NS_Zak}
    Z^\lambda = -\oint_{\mathtt{BZ}}dk \langle \widetilde{\psi}_\lambda(k)|\partial_k{ \widetilde{\psi}_\lambda (k)} \rangle,
\end{equation}
where $|\widetilde{\psi}_\lambda(k)\rangle$ is defined as
\begin{equation}\label{eq:NS_def}
    |\widetilde{\psi}_\lambda(k)\rangle = \frac{\langle A|\psi_\lambda(k)\rangle}{
    \sqrt{\langle \psi_{\lambda}(k)|A\rangle\langle A|\psi_{\lambda}(k)\rangle}}|A\rangle. 
\end{equation}
Here $|A\rangle$ denotes the unit vector of the first sublattice. This definition corresponds to projecting the eigenstate onto the first sublattice, followed by normalization. To obtain the non-Hermitian version of NS Zak's phase, the first step is to replace all $\langle \psi_\lambda(k)|$ with left vectors $\langle \phi_\lambda(k)|$ in Eq.~(\ref{eq:NS_def}), which leads to
\begin{equation}
\begin{aligned}
    |\widetilde{\psi}_\lambda(k)\rangle_{\mathtt{NH}} &= \frac{\langle A|\psi_\lambda(k)\rangle}{
    \sqrt{\langle \phi_{\lambda}(k)|A\rangle\langle A|\psi_{\lambda}(k)\rangle}}|A\rangle \\
    &= \sqrt{\frac{a^\lambda_R}{a^\lambda_L}}e^{\frac{i(\theta^\lambda_L+\theta^\lambda_R)}{2}}|A\rangle,
\end{aligned}
\end{equation}
where $a^\lambda_{L/R}$ and $\theta^\lambda_{L/R}$ denote the modulus and argument phases of the projected left/right vector of band $\lambda$ on sublattice A. Similarly, we have
\begin{equation}
    |\widetilde{\phi}_\lambda(k)\rangle_{\mathtt{NH}} = \sqrt{\frac{a^\lambda_L}{a^\lambda_R}}e^{\frac{i(\theta^\lambda_L+\theta^\lambda_R)}{2}}|A\rangle.
\end{equation}
Thus, the NS Zak's phase in non-Hermitian SSH3 model can be expressed as 
\begin{equation}
\begin{aligned}
    Z^\lambda_{\mathtt{NH}} &= -\oint_{\mathtt{GBZ}}dk \langle \widetilde{\phi}_\lambda(k)|\partial_k{ \widetilde{\psi}_\lambda (k)} \rangle\\
    &= -\frac{1}{2}\oint_{\mathtt{GBZ}}dk (\frac{\partial \theta^\lambda_L}{\partial k} + \frac{\partial \theta^\lambda_R}{\partial k}) - \oint_{\mathtt{GBZ}}d\mathtt{ln}\sqrt{\frac{a^\lambda_R}{a^\lambda_L}}\\
    &= -\frac{1}{2}\oint_{\mathtt{GBZ}}dk (\frac{\partial \theta^\lambda_L}{\partial k} + \frac{\partial \theta^\lambda_R}{\partial k}).
\end{aligned}
\end{equation}
Here the second term vanishes upon loop integration due to the single-valued nature of the modulus. This result can be interpreted as the mean value of the cumulative phase after a loop of the left and right vectors.

In the next step, we show that the cumulative phase for the left and right vectors are the same. Similar to Eq.~(\ref{eq:H(beta)_transform}) in Sec.~\ref{sec:GBZ}, $H(r_M^2/\beta^*)$ and $H^\dagger(\beta)$ is also connected via a similarity transformation $S_{\eta}$, namely
\begin{equation}
    S_{\eta}^{-1}H(r_M^2/\beta^*)S_{\eta} = H^\dagger(\beta).
\end{equation}
Since the GBZ is a circle with radius $|r_M|$ and can be parameterized as $\beta = |r_M|e^{ik}$, we have
\begin{equation}
    S_{\eta}^{-1}H(\beta(k))S_{\eta} = H^\dagger(\beta(k))
\end{equation}
when $r_M^2 > 0$ and
\begin{equation}
    S_{\eta}^{-1}H(\beta(k+\pi))S_{\eta} = H^\dagger(\beta(k))
\end{equation}
when $r_M^2 < 0$. For the first case where $r_M^2 > 0$, $|\psi_\lambda(k)\rangle$ and $S_\eta|\phi_\lambda(k)\rangle$ are linearly dependent due to the similarity transformation. The exact proportional ratio between these two vectors is irrelevant as we only care about the relative phase between the first and third sublattices. Recall that the expression for $S_\eta$ reads
\begin{equation}
    S_\eta = \mathtt{diag}\{1, \frac{t_\mathtt{R_1}}{t_\mathtt{L_1}},\frac{t_\mathtt{R_1}t_\mathtt{R_2}}{t_\mathtt{L_1}t_\mathtt{L_2}}\},
\end{equation}
we learn that for $(t_\mathtt{R_1} t_\mathtt{R_2})/(t_\mathtt{L_1} t_\mathtt{L_2}) > 0$ the relative phase between the first and third sublattices is the same for $|\psi_\lambda(\beta(k))\rangle$ and $|\phi_\lambda(\beta(k))\rangle$; and for $(t_{\mathtt{R_1}}t_{\mathtt{R_2}})/(t_{\mathtt{L}_1}t_{\mathtt{L}_2} )< 0$, the relative phase is different by $\pi$ for $|\psi_\lambda(\beta(k))\rangle$ and $|\phi_\lambda(\beta(k))\rangle$. Note that $(t_\mathtt{R_1}t_\mathtt{R_2}/t_\mathtt{L_1}t_\mathtt{L_2})$ is constant, so we always have
\begin{equation}
    \frac{\partial \theta^\lambda_L}{\partial k} = \frac{\partial \theta^\lambda_R}{\partial k}.
\end{equation}
For the $r_M^2 < 0$ case,  $|\psi_\lambda(\beta(k+\pi))\rangle$ and $S_\eta|\phi_\lambda(\beta(k))\rangle$ are linearly dependent. Similarly, we have
\begin{equation}
    \frac{\partial \theta^\lambda_L(k)}{\partial k} = \frac{\partial \theta^\lambda_R(k+\pi)}{\partial k}.
\end{equation}
The loop integral in GBZ is equal for $\theta_L^\lambda$ and $\theta_R^\lambda$, owing to their periodicity in $k$. Thus, we prove that the contribution of both left and right vectors to the NS Zak's phase is identical. Hence, the expression in the Hermitian case can be safely used with BZ replaced by GBZ, namely
\begin{equation}
    Z^\lambda_{\mathtt{NH}} = -\oint_{\mathtt{GBZ}}dk\frac{\partial \theta_R^\lambda}{\partial k}.
\end{equation}

In the end, we explain the axes in Fig.~\ref{fig:5}(a). By choosing $(t_\mathtt{L_1}t_\mathtt{R_1})/t_3^2$ and $(t_\mathtt{L_2}t_\mathtt{R_2})/t_3^2$ as the axes, we indicate that the result is invariant for different parameter sets as long as these two quantities are invariant. First, we show straightforwardly that the OBC spectrum should be invariant. Consider two different OBC Hamiltonians $H_1, H_2$, the only difference is that we have $t_\mathtt{L_1}, t_\mathtt{R_1}$ in the first one and $t_\mathtt{L_1}^\prime=\alpha t_\mathtt{L_1}, t_\mathtt{R_1}^\prime=t_\mathtt{R_1}/\alpha$ 
for $\alpha \in \mathbb{R}$ in the second one, so that $t_\mathtt{L_1}t_\mathtt{R_1}=t_\mathtt{L_1}^\prime t_\mathtt{R_1}^\prime$. The IGT
\begin{equation}
    S = \mathtt{diag}\{\alpha^{-1}, \alpha^{-2}, \cdots, \alpha^{-N}\} \otimes \mathtt{diag}\{1, \alpha^{-1}, \alpha^{-1}\}
\end{equation}
can transform $H_2$ to $H_1$, or symbolically $S^{-1}H_2S=H_1$. Hence, the energy spectrum for both continuum band and discrete levels are the same for these two parameter sets. 

Then, we show that the NS Zak's phase should also produce the same result. By applying the similarity transformation 
\begin{equation}
    S^{\prime} = \mathtt{diag}\{1, \alpha^{-1}, \alpha^{-1}\}
\end{equation}
to the generalized Bloch Hamiltonian $H_2(\beta)$, we have
\begin{equation}
    S^{\prime -1}H_2(\beta)S^{\prime} = H_1(\beta\alpha).
\end{equation}
Note that the radius of GBZ for $H_2$ is $r/|\alpha|$ if that for $H_1$ is $r$, we have for any $\beta$ in the GBZ of $H_2$, $\alpha\beta$ is in the GBZ of $H_1$. Thus, the eigenstates in GBZ of $H_1$ and $H_2$ are connected by this similarity transformation, and by following the same analysis, we obtain that the NS Zak's phase for two cases is the same.

\bibliography{mybib}

\end{document}